# Privatization-Safe Transactional Memories (Extended Version)

**Artem Khyzha**
Tel Aviv University, Tel Aviv, Israel

**Hagit Attiya**
Technion — Israel Institute of Technology, Haifa, Israel

**Alexey Gotsman**
IMDEA Software Institute, Madrid, Spain

## Abstract

Transactional memory (TM) facilitates the development of concurrent applications by letting the programmer designate certain code blocks as atomic. Programmers using a TM often would like to access the same data both inside and outside transactions, and would prefer their programs to have a *strongly atomic* semantics, which allows transactions to be viewed as executing atomically with respect to non-transactional accesses. Since guaranteeing such semantics for arbitrary programs is prohibitively expensive, researchers have suggested guaranteeing it only for certain *data-race free (DRF)* programs, particularly those that follow the *privatization* idiom: from some point on, threads agree that a given object can be accessed non-transactionally.

In this paper we show that a variant of *Transactional DRF (TDRF)* by Dalessandro et al. is appropriate for a class of *privatization-safe* TMs, which allow using privatization idioms. We prove that, if such a TM satisfies a condition we call *privatization-safe opacity* and a program using the TM is TDRF under strongly atomic semantics, then the program indeed has such semantics. We also present a method for proving privatization-safe opacity that reduces proving this generalization to proving the usual opacity, and apply the method to a TM based on two-phase locking and a privatization-safe version of TL2. Finally, we establish the inherent cost of privatization-safety: we prove that a TM cannot be progressive and have invisible reads if it guarantees strongly atomic semantics for TDRF programs.

**2012 ACM Subject Classification** Theory of computation → Concurrency; Theory of computation → Program semantics; Software and its engineering → Software verification

**Digital Object Identifier** 10.4230/LIPIcs.DISC.2019.25

**Funding** This research was funded in part by the Israel Science Foundation (grants 2005/17, 1749/14 and 380/18) and the European Research Council (Starting Grant RACCOON).

## 1 Introduction

*Transactional memory* (TM) facilitates the development of concurrent applications by letting the programmer designate certain code blocks as *atomic* [23]. TM allows developing a program and reasoning about its correctness as if each atomic block executes as a *transaction*—atomically and without interleaving with other blocks—even though in reality the blocks can be executed concurrently. A TM can be implemented in hardware [24, 28], software [33] or a combination of both [13, 27].

Often programmers using a TM would like to access the same data both inside and outside transactions. This may be desirable to avoid performance overheads of transactional accesses, to support legacy code, or for explicit memory deallocation. One typical pattern is *privatization* [30, 34], illustrated in Figure 1. There the `atomic` blocks return a value signifying whether the transaction committed or aborted. In the program, an object `x` is

```
{ priv = false ∧ x = 0 }
l1 = atomic {           || atomic {
  priv = true; } // T1  ||   if (!priv)
if (l1 == committed)    ||     x = 42;
  x = 1; // n           || } // T2
{ l1 = committed ⟹ x = 1 }
```

**Figure 1** Privatization.

```
{ priv = true ∧ x = l = 0 }
x = 42; // n            || l2 = atomic {
l1 = atomic {           ||   if (!priv)
  priv = false;         ||     l = x;
} // T1                 || } // T2
{ l2 = committed ∧ l ≠ 0 ⟹ l = 42 }
```

**Figure 2** Publication.

```
{ x = y = 0 }
atomic {    ||
  x = 1;    || l1 = x; // n1
  y = 2;    || l2 = y; // n2
} // T      ||
{ l1 = 1 ⟹ l2 = 2 }
```

**Figure 3** Data race.

guarded by a flag `priv`, showing whether the object should be accessed transactionally (`false`) or non-transactionally (`true`). The left-hand-side thread first tries to set the flag inside transaction $T_1$, thereby *privatizing* `x`. If successful, it then accesses `x` non-transactionally. A concurrent transaction $T_2$ in the right-hand-side thread checks the flag `priv` prior to accessing `x`, to avoid simultaneous transactional and non-transactional access to the object. We expect the postcondition shown to hold: if privatization is successful, at the end of the program `x` should store 1, not 42. The opposite idiom is *publication*, illustrated in Figure 2. The left-hand-side thread writes to `x` non-transactionally and then clears the flag `priv` inside transaction $T_1$, thereby *publishing* `x`. The right-hand-side thread tests the flag inside transaction $T_2$, and if it is cleared, reads `x`. Again, we expect the postcondition to hold: if the right-hand-side thread sees the write to the flag, it should also see the write to `x`. The two idioms can be combined: the programmer may privatize an object, then access it non-transactionally, and finally publish it back for transactional access.

Ideally, programmers mixing transactional and non-transactional accesses to objects would like their programs to have *strongly atomic semantics* [8], where transactions can be viewed as executing atomically also with respect to non-transactional accesses, i.e., without interleaving with them. This is equivalent to considering every non-transactional access as a single-instruction transaction. For example, the program in Figure 3 under strongly atomic semantics can only produce executions where each of the non-transactional accesses $n_1$ and $n_2$ executes either before or after the transaction $T$, so that the postcondition in Figure 3 always holds. Unfortunately, providing such semantics in software requires instrumenting non-transactional accesses with additional instructions for maintaining TM metadata [19]. This undermines scalability and makes it difficult to reuse legacy code. Since most existing TMs are either software-based or rely on a software fall-back, they do not perform such instrumentation and, hence, provide weaker atomicity guarantees. For example, they may allow the program in Figure 3 to execute non-transactional accesses $n_1$ and $n_2$ between transactional writes to `x` and `y` and, thus, observe an intermediate state of the transaction, e.g., $\mathtt{x} = 1$ and $\mathtt{y} = 0$, violating the postcondition in Figure 3.

Researchers have suggested resolving the tension between strong TM semantics and performance by guaranteeing strongly atomic semantics only to *data-race free (DRF)* programs—informally, programs without concurrent transactional and non-transactional accesses to the same data [4, 5, 10, 11, 30, 32, 34]. For example, we do not have to guarantee strongly atomic semantics for the program in Figure 3, which has such concurrent accesses to `x` and `y`. On the other hand, the programs in Figure 1 and Figure 2 should be guaranteed strongly atomic semantics, since at any point of time, an object is accessed either only transactionally or only non-transactionally. Despite the intuitive simplicity of this idea, coming up with a precise DRF definition is nontrivial: early on there were multiple competing proposals for the notion of DRF, and it was unclear how to select among them [4, 10, 11, 25, 29]. To address this, we have recently formalized the requirements on an appropriate notion of DRF using observational refinement [26]: a TM needs to guarantee that, if a program is DRF under the strongly atomic semantics (formalized as *transactional sequential consistency* [11]), then all its executions are observationally equivalent to strongly atomic ones. This *Fundamental Property* allows the programmer to never reason about weakly atomic semantics at all, even when checking DRF.

Different TMs have different requirements on mixing transactional and non-transactional accesses needed to validate the Fundamental Property. *Privatization-safe* TMs, such as lock-based TMs [15, 21] and NOrec [12], allow the programmer to ensure the absence of concurrent transactional and non-transactional accesses by synchronizing them using transactional operations. Then the program in Figure 1, which synchronizes accesses to `x` using `priv`, is guaranteed strongly atomic semantics as is. *Privatization-unsafe* TMs, such as TL2 [14] and TinySTM [16], require the programmer to insert additional synchronization, e.g., via *transactional fences* [30, 34], which block until all the transactions that were active when the fence was invoked complete. For example, such TMs do not guarantee strongly atomic semantics to the program in Figure 1 unless the transaction $T_1$ is immediately followed by a transactional fence. This is because TMs such as TL2 execute transactions optimistically, flushing their writes to memory only on commit. Then, in the absence of a fence, the transaction $T_1$ can privatize `x` and $n$ can modify it after $T_2$ started committing, but before its write to `x` reached the memory, so that $T_2$'s write subsequently overwrites $n$'s write and violates the postcondition. TMs that make transactional updates in-place and undo them on abort are subject to a similar problem.

Privatization-safe TMs provide a simpler programming model, since they do not require the programmer to select where to place fences. However, the programmer still needs to avoid programs of the kind shown in Figure 3, which would lead the TM to violate strong atomicity. In this paper we show that a variant of *transactional DRF (TDRF)* previously proposed by Dalessandro et al. [11] is appropriate to formalize the programmer's obligations. To this end, we show that this variant of TDRF validates the Fundamental Property, provided the TM satisfies a generalization of *opacity* [20, 21], which we call *privatization-safe opacity*. To formulate this kind of opacity, we generalize TDRF to arbitrary TM histories, not just strongly atomic ones. These results complement our previous proposal of DRF for privatization-unsafe TMs, which considers a more low-level programming model requiring fence placements [26].

We furthermore present a method for proving privatization-safe opacity and apply it to a TM based on two-phase locking [21] and a privatization-safe version of TL2 [14] that executes a fence at the end of each transaction. A key feature of our method is that it reduces proving privatization-safe opacity to proving the ordinary opacity of the TM assuming no mixed transactional/non-transactional accesses. This allows us to reuse the previous proofs of opacity of the two-phase locking TM [21] and TL2 [26].

Finally, our framework allows proving an interesting result about the inherent cost of privatization-safety. We prove that a TM that provides strongly atomic semantics to TDRF programs cannot be *progressive* and have *invisible reads*: it cannot ensure that transactions always complete when running solo and also that transactions reading objects do not prevent transactions writing to them from committing. This result significantly simplifies and strengthens a lower bound by Attiya and Hillel [7], which did not use a formal DRF notion.

## 2 Programming Language and Strongly Atomic Semantics

**Language syntax.** We formalize our results for a simple programming language with mixed transactional and non-transactional accesses. A *program* $P = C_1 \parallel \ldots \parallel C_N$ in our language is a parallel composition of *commands* $C_t$ executed by different *threads* $t \in \mathsf{ThreadID} =$ $\{1, \ldots, N\}$. Every thread $t \in \mathsf{ThreadID}$ has a set of *local variables* $l \in \mathsf{LVar}_t$, which only it can access; for simplicity, we assume that these are integer-valued. Threads have access to a *transactional memory* (*TM*), which manages a fixed collection of *shared register objects* $x \in \mathsf{Reg}$. The syntax of commands $C \in \mathsf{Com}$ is as follows:

$$C ::= c \mid C\,;C \mid \ \mathtt{if}\,(b)\ \mathtt{then}\ C\ \mathtt{else}\ C \mid \ \mathtt{while}\,(b)\ \mathtt{do}\ C$$
$$\mid l = \mathtt{atomic}\,\{C\} \mid l = x.\mathtt{read}() \mid x.\mathtt{write}(e)$$

where $b$ and $e$ denote Boolean, respectively, integer *expressions* over local variables and constants. The language includes *primitive commands* $c \in \mathsf{PCom}$, which operate on local variables, and standard control-flow constructs. An *atomic block* $l = \mathtt{atomic}\,\{C\}$ executes $C$ as a *transaction*, which the TM can *commit* or *abort*. The system's decision is returned in the local variable $l$, which receives a distinguished value committed or aborted. We do not allow programs to abort a transaction explicitly and forbid nested atomic blocks. Threads can invoke two methods on a register $x$: $x.\mathtt{read}()$ returns the current value of $x$, and $x.\mathtt{write}(e)$ sets it to $e$. These methods may be invoked both *inside* and *outside* atomic blocks.

**Model of computations.** The semantics of our programming language is defined in terms of *traces*—certain finite sequences of *actions*, each describing a single computation step (in this paper we consider only finite computations). Let ActionId be a set of *action identifiers*. Actions are of two kinds. A *primitive action* denotes the execution of a primitive command and is of the form $(a, t, c)$, where $a \in \mathsf{ActionId}$, $t \in \mathsf{ThreadID}$ and $c \in \mathsf{PCom}$. An *interface action* has one of the following forms (where $x \in \mathsf{Reg}$ and $v \in \mathbb{Z}$):

| Request actions | Matching response actions |
|---|---|
| $(a, t, \mathsf{begintx})$ | $(a, t, \mathsf{ok}) \mid (a, t, \mathsf{aborted})$ |
| $(a, t, \mathsf{trycommit})$ | $(a, t, \mathsf{committed}) \mid (a, t, \mathsf{aborted})$ |
| $(a, t, \mathsf{write}(x, v))$ | $(a, t, \mathsf{ret}(\bot)) \mid (a, t, \mathsf{aborted})$ |
| $(a, t, \mathsf{read}(x))$ | $(a, t, \mathsf{ret}(v)) \mid (a, t, \mathsf{aborted})$ |

Interface actions usually denote the control flow of a thread $t$ crossing the boundary between the program and the TM: *request* actions correspond to the control being transferred from the former to the latter, and *response* actions, the other way around. A begintx action is generated upon entering an `atomic` block, and a trycommit action when a transaction tries to commit upon exiting an `atomic` block. The request actions $\mathsf{write}(x, v)$ and $\mathsf{read}(x)$ denote invocations of the `write`, respectively, `read` methods of register $x$; a `write` action is annotated with the value $v$ written. The response actions $\mathsf{ret}(\bot)$ and $\mathsf{ret}(v)$ denote the return from invocations of `write`, respectively, `read` methods of a register; the latter is annotated with the value $v$ read. The TM may abort a transaction at any point when it is in control; this is recorded by an aborted response action. To simplify notation, we reuse the interface actions for reads and writes to denote accesses outside transactions.

A *trace* $\tau$ is a finite sequence of actions satisfying the expected well-formedness conditions, e.g., that request and response actions are properly matched, and so are actions denoting the beginning and the end of transactions (we defer the formal definition to §A). A *transaction* $T$ is a nonempty trace such that it contains actions by the same thread, begins with a begintx action and only its last action can be a committed or an aborted action. A transaction $T$ is: *committed* if it ends with a committed action, *aborted* if it ends with aborted, *commit-pending* if it ends with trycommit, and *live*, in all other cases. A transaction $T$ is in a trace $\tau$ if $T$ is a subsequence of $\tau$ and no longer transaction is. We refer to interface actions in a trace outside of a transaction as *non-transactional actions*. We call a matching request/response pair of a read or a write a *non-transactional access* (ranged over by $n$).

A *history* is a trace containing only interface actions (thus, omitting all accesses to local variables); we use $H, S$ to range over histories, and $H(i)$ to refer to the $i$-th action in $H$. We also use $\mathsf{history}(\tau)$ to denote a projection of a trace to interface actions. Since histories fully capture the possible interactions between a TM and a client program, we often conflate the notion of a TM and the set of histories it produces. Hence, a *transactional memory* $\mathcal{H}$ is a prefix-closed set of histories. We assume that a TM always allows a client program to execute a request and, hence, require $\mathcal{H}$ to be closed under appending any request action to its histories, provided that the latter remain well-formed. Note that histories include actions corresponding to non-transactional accesses, even though these may not be directly managed by the TM implementation. This is needed to account for changes to registers performed by such actions when defining the TM semantics: e.g., in the case when a register is privatized, modified non-transactionally and then published back for transactional access. Of course, a well-formed TM semantics should not impose restrictions on the placement of non-transactional actions, since these are under the control of the program.

**Strongly atomic semantics.** The *semantics* of a program $P$ is given by the set $[\![P]\!](\mathcal{H})$ of traces it produces when executed with a TM $\mathcal{H}$. Its formal definition follows the intuitive meaning of commands, and we defer it to §A. Our semantics assumes that the underlying memory is sequentially consistent, which allows us to focus on the key issues specific to TM (we leave handling weak memory for future work, discussed in §9). We use the semantics instantiated with one particular TM to define the *strongly atomic* semantics of programs [8], which is equivalent to transactional sequential consistency [11]. Following [6], we use an *atomic* TM $\mathcal{H}_{\mathsf{atomic}}$ for this purpose: the strongly atomic semantics of a program $P$ is given by the set of traces $[\![P]\!](\mathcal{H}_{\mathsf{atomic}})$. The TM $\mathcal{H}_{\mathsf{atomic}}$ contains only histories that are *non-interleaved*, i.e., where actions by one transaction do not overlap with actions of another transaction or of non-transactional accesses. Out of such histories, $\mathcal{H}_{\mathsf{atomic}}$ contains only histories following the intuitive atomic semantics of transactions: every response action of a $\mathsf{read}(x)$ returns the value $v$ in the last preceding $\mathsf{write}(x, v)$ action that is not located in an aborted or live transaction different from the one of the $\mathsf{read}$; if there is no such $\mathsf{write}$, the read returns the initial value $v_{\mathsf{init}}$. We defer a formal definition of $\mathcal{H}_{\mathsf{atomic}}$ to §A.

## 3 Transactional Data-Race Freedom

We now formalize in our framework a variant of *transactional data-race freedom (TDRF)* of Dalessandro et al. [11]. According to this notion, a data race happens between a pair of *conflicting* actions, as defined below.

▶ **Definition 1.** *A non-transactional request action $\alpha$ and a transactional request action $\alpha'$* conflict *if $\alpha$ and $\alpha'$ are executed by different threads, they are read or write actions on the same register, and at least one of them is a write.*

As is standard, we formalize when conflicting actions form a data race using a *happens-before* relation $\mathsf{hb}(H)$ on actions in a history $H$. We first define the *execution order* of $H$ as follows: $\alpha <_H \alpha'$ iff for some $i$ and $j$, $\alpha = H(i)$, $\alpha' = H(j)$ and $i < j$.

▶ **Definition 2.** *The* happens-before *relation of a history $H \in \mathcal{H}_{\mathsf{atomic}}$ is* $\mathsf{hb}(H) \triangleq (\mathsf{po}(H) \cup \mathsf{ef}(H) \cup \mathsf{cl}(H))^+$, *where*

- per-thread order $\mathsf{po}(H)$: *$\alpha <_{\mathsf{po}(H)} \alpha'$ iff $\alpha <_H \alpha'$ and $\alpha, \alpha'$ are by the same thread;*
- effect order $\mathsf{ef}(H)$: *$\alpha <_{\mathsf{ef}(H)} \alpha'$ iff $\alpha <_H \alpha'$ and $\alpha, \alpha'$ are by different transactions;*
- client order $\mathsf{cl}(H)$: *$\alpha <_{\mathsf{cl}(H)} \alpha'$ iff $\alpha <_H \alpha'$ and $\alpha, \alpha'$ are non-transactional in $H$.*

▶ **Definition 3.** *A history* $H \in \mathcal{H}_{\mathsf{atomic}}$ *is* transactional data-race free*, written* $\mathsf{TDRF}(H)$*, if every pair of conflicting actions in it is ordered by* $\mathsf{hb}(H)$ *one way or another. A program* $P$ *is* transactional data-race free*, written* $\mathsf{TDRF}(P)$*, if* $\forall \tau \in [\![P]\!](\mathcal{H}_{\mathsf{atomic}}).\, \mathsf{TDRF}(\mathsf{history}(\tau))$.

Components of happens-before used to define TDRF describe various forms of synchronization available in our programming language. First, actions by the same thread cannot be concurrent and thus we let $\mathsf{po}(H) \subseteq \mathsf{hb}(H)$. Second, privatization-safe TMs provide synchronization between transactions, which follows their order in non-interleaved histories of an atomic TM considered in the definition of TDRF on programs. Thus, we let $\mathsf{ef}(H) \subseteq \mathsf{hb}(H)$. Finally, we let $\mathsf{cl}(H) \subseteq \mathsf{hb}(H)$, because in this paper we assume a sequentially consistent memory model and, hence, do not consider pairs of conflicting non-transactional accesses as races. This is the key difference between our variant of TDRF and the original definition by Dalessandro et al. [11], which does not include the client order into happens-before. Our variant of TDRF imposes fewer obligations on the programmer: as we show by establishing the Fundamental Property for our variant of TDRF (§5), under sequentially consistent memory races on non-transactional accesses are harmless for privatization-safety.

To illustrate the TDRF definition, we show that the program in Figure 1 is TDRF by considering the histories it produces with the atomic TM (the program in Figure 2 can be shown TDRF analogously). The possible conflicts are between the accesses to `x` in $n$ and $T_2$. For a conflict to occur, $T_2$ has to read `false` from `priv`; then $T_2$ has to execute before $T_1$, yielding a history of the form $T_2 T_1 n$. In this history $T_2$ precedes $T_1$ in the effect order and $T_1$ precedes $n$ in the per-thread order, meaning that $\mathsf{hb}(H)$ orders the conflict between $T_2$ and $n$. Similarly, in §B we show that programs following a *proxy privatization* pattern [36], where one thread privatizes an object for another thread, are also TDRF. On the other hand, the program in Figure 3 is not TDRF, since in histories it produces with the atomic TM, the happens-before never relates $T$ with $n_1$ and $n_2$. Finally, the inclusion of $\mathsf{cl}(H) \subseteq \mathsf{hb}(H)$ allows us to consider DRF those programs that privatize an object by agreeing on its status outside transactions ("partitioning by consensus" in [34]); we provide an example in §B.

## 4 Privatization-Safe Opacity

We now present our first contribution—a generalization of *opacity* of a TM $\mathcal{H}$ [20, 21] that guarantees that the TM provides strongly atomic semantics to TDRF programs. We call this generalization *privatization-safe opacity*. Its definition requires that a history $H$ of a TM $\mathcal{H}$ can be matched by a history $S$ of the atomic TM $\mathcal{H}_{\mathsf{atomic}}$ that "looks similar" to $H$ from the perspective of the program. The similarity is formalized by a relation $H \sqsubseteq S$, which requires $S$ to be a permutation of $H$ preserving its per-thread and client orders.

▶ **Definition 4.** *A history* $H_1$ *corresponds to a history* $H_2$*, written* $H_1 \sqsubseteq H_2$*, if there is a bijection* $\theta : \{1, \ldots, |H_1|\} \to \{1, \ldots, |H_2|\}$ *such that* $\forall i.\, H_1(i) = H_2(\theta(i))$ *and*

$$\forall i, j.\, i < j \wedge H_1(i) <_{\mathsf{po}(H_1) \cup \mathsf{cl}(H_1)} H_2(j) \implies \theta(i) < \theta(j).$$

The above relation differs in several ways from the one used to define the ordinary opacity. First, unlike in the ordinary opacity, our histories include non-transactional actions, because these can affect the behavior of the TM. Second, instead of preserving $\mathsf{cl}(H_1)$ in Definition 4, the ordinary opacity requires preserving the following *real-time order* $\mathsf{rt}(H_1)$ on actions: $\alpha <_{\mathsf{rt}(H)} \alpha'$ iff $\alpha \in \{(_, _, \mathsf{committed}), (_, _, \mathsf{aborted})\}$, $\alpha' = (_, _, \mathsf{begintx})$ and $\alpha <_H \alpha'$. This orders non-overlapping transactions, with the duration of a transaction determined by the interval from its $\mathsf{begintx}$ action to the corresponding $\mathsf{committed}$ or $\mathsf{aborted}$ action (or to the end of the history if there is none). However, preserving real-time order is unnecessary if all means of communication between program threads are reflected in histories [17].

We next lift privatization-safe opacity to TMs. A straightforward definition, mirroring the ordinary opacity, would require any history of the TM $\mathcal{H}$ to have a matching history of the atomic TM $\mathcal{H}_{\mathsf{atomic}}$. However, such a requirement would be too strong for our setting: since the TM has no control over non-transactional actions of its clients, histories in $\mathcal{H}$ may be produced by racy programs, and we do not want to require the TM to guarantee strong atomicity in such cases. For example, even though a simple TM based on a single global lock is privatization-safe, it has a history produced by the program from Figure 3 that does not have a matching history of $\mathcal{H}_{\mathsf{atomic}}$ (§1). Hence, our definition of privatization-safe opacity requires only histories produced by TDRF programs to have justifications in $\mathcal{H}_{\mathsf{atomic}}$. To express this restriction, we generalize data-race freedom to be defined over an arbitrary concurrent history $H$, not just one produced by $\mathcal{H}_{\mathsf{atomic}}$. The new DRF requires that every history of the atomic TM matching $H$ according to the opacity relation be TDRF.

▶ **Definition 5.** *A history $H \in \mathcal{H}$ is* concurrent data-race free*, written* $\mathsf{CDRF}(H)$*, if* $\forall S \in \mathcal{H}_{\mathsf{atomic}}.\, H \sqsubseteq S \implies \mathsf{TDRF}(S)$*. Let* $\mathcal{H}|_{\mathsf{CDRF}} = \{H \in \mathcal{H} \mid \mathsf{CDRF}(H)\}$*. A program $P$ is concurrent data-race free with a TM $\mathcal{H}$, written* $\mathsf{CDRF}(P, \mathcal{H})$*, if* $\forall \tau \in [\![P]\!](\mathcal{H}).\, \mathsf{CDRF}(\mathsf{history}(\tau))$*.*

▶ **Definition 6.** *A TM $\mathcal{H}$ is* privatization-safe opaque*, written* $\mathcal{H}|_{\mathsf{CDRF}} \sqsubseteq \mathcal{H}_{\mathsf{atomic}}$*, if for every history $H \in \mathcal{H}|_{\mathsf{CDRF}}$ there exists a history $S \in \mathcal{H}_{\mathsf{atomic}}$ such that $H \sqsubseteq S$ holds.*

The following lemma (proved in §C) justifies using CDRF as a generalization of TDRF to concurrent histories by establishing that TDRF programs indeed produce CDRF histories.

▶ **Lemma 7.** *For every program $P$ and a TM system $\mathcal{H}$,* $\mathsf{TDRF}(P)$ *implies* $\mathsf{CDRF}(P, \mathcal{H})$*.*

## 5 The Fundamental Property

We next formalize the Fundamental Property of TDRF using *observational refinement* [6]: if a program is TDRF under the atomic TM $\mathcal{H}_{\mathsf{atomic}}$, then any trace of the program under a privatization-safe opaque TM $\mathcal{H}$ has an *observationally equivalent* trace under $\mathcal{H}_{\mathsf{atomic}}$.

▶ **Definition 8.** *Traces $\tau$ and $\tau'$ are* observationally equivalent*, denoted by $\tau \sim \tau'$, if $\forall t.\, \tau|_t = \tau'|_t$ and $\tau|_{\mathsf{nontx}} = \tau'|_{\mathsf{nontx}}$, where $\tau|_{\mathsf{nontx}}$ denotes the subsequence of $\tau$ containing all actions from non-transactional accesses.*

Equivalent traces are considered indistinguishable to the user. In particular, the sequences of non-transactional accesses in equivalent traces (which usually include all I/O) satisfy the same linear-time temporal properties. We lift the equivalence to sets of traces as follows.

▶ **Definition 9.** *A set of traces $\mathcal{T}$* observationally refines *a set of traces $\mathcal{T}'$, written $\mathcal{T} \preceq \mathcal{T}'$, if $\forall \tau \in \mathcal{T}.\, \exists \tau' \in \mathcal{T}'.\, \tau \sim \tau'$.*

▶ **Theorem 10** (Fundamental Property)**.** *If $\mathcal{H}$ is a TM such that $\mathcal{H}|_{\mathsf{CDRF}} \sqsubseteq \mathcal{H}_{\mathsf{atomic}}$ and $P$ is a program such that* $\mathsf{TDRF}(P)$*, then* $[\![P]\!](\mathcal{H}) \preceq [\![P]\!](\mathcal{H}_{\mathsf{atomic}})$*.*

Theorem 10 establishes a contract between the programmer and the TM implementors. The TM implementor has to ensure privatization-safe opacity of the TM assuming the program is DRF: $\mathcal{H}|_{\mathsf{CDRF}} \sqsubseteq \mathcal{H}_{\mathsf{atomic}}$. The programmer has to ensure the DRF of the program under strongly atomic semantics: $\mathsf{TDRF}(P)$. This contract lets the programmer check properties of a program assuming strongly atomic semantics ($[\![P]\!](\mathcal{H}_{\mathsf{atomic}})$) and get the guarantee that the properties hold when the program uses the actual TM implementation ($[\![P]\!](\mathcal{H})$). Theorem 10 follows from Lemma 7 and the next lemma, which is an adaptation of a result from [6].

▶ **Lemma 11.** *If* $\mathcal{H}$ *is a TM such that* $\mathcal{H}|_{\mathsf{CDRF}} \sqsubseteq \mathcal{H}_{\mathsf{atomic}}$*, then* $\forall P.\, \mathsf{CDRF}(P, \mathcal{H}) \implies [\![P]\!](\mathcal{H}) \preceq [\![P]\!](\mathcal{H}_{\mathsf{atomic}})$*.*

## 6 Proving Privatization-Safe Opacity

We now develop a method that reduces proving privatization-safe opacity ($\mathcal{H}|_{\mathsf{CDRF}} \sqsubseteq \mathcal{H}_{\mathsf{atomic}}$) to proving the ordinary opacity. The method builds on a *graph characterization* of opacity by Guerraoui and Kapalka [21], which was proposed for proving opacity of TMs that do not allow mixed transactional/non-transactional accesses to the same data. The characterization allows checking opacity of a history $H$ by checking two properties: *consistency* of the history, denoted $\mathsf{cons}(H)$, and the acyclicity of a certain *opacity graph*, which we define in the following. Consistency is a basic well-formedness property of a history ensuring the following. If a transaction $T$ in $H$ reads a value of a register $x$ and writes to it before, then $T$ reads the latest value it writes. If $T$ reads a value of $x$ and does not write to it before, then it reads some value written non-transactionally or by a committed or commit-pending transaction (or the initial value, when everything else fails). Consistency also ensures that only the last write to $x$ by a transaction is read from. We define consistency formally in §D and focus here on defining opacity graphs.

The vertexes in these graphs include transactions and non-transactional accesses in $H$. The intention of the vis predicate below is to mark those vertexes that have taken effect, including commit-pending transactions of this kind. The other components, intuitively, constrain the order in which the vertexes should appear in the atomic history.

▶ **Definition 12.** *The* opacity graph *of a history* $H$ *is a tuple* $G = (\mathcal{V}, \mathsf{vis}, \mathsf{WR}, \mathsf{WW}, \mathsf{RW}, \mathsf{PO}, \mathsf{CL})$*, where:*

- $\mathcal{V}$ *is the set of graph vertexes, i.e., all transactions and non-transactional accesses from* $H$ *(ranged over by* $\nu$*).*
- $\mathsf{vis} \subseteq \mathcal{V}$ *is a* visibility *predicate, which holds of all non-transactional accesses and committed transactions and does not hold of all aborted and live transactions.*
- $\mathsf{WR} : \mathsf{Reg} \to 2^{\mathcal{V} \times \mathcal{V}}$ *specifies per-register* read-dependency *relations on vertexes, such that*
  - *For each read dependency* $\nu \xrightarrow{\mathsf{WR}_x} \nu'$*, we have that* $\nu \neq \nu'$*,* $\nu$ *contains* $(_, _, \mathsf{write}(x, v))$*, and* $\nu'$ *contains a request* $(_, _, \mathsf{read}(x))$ *and a matching response* $(_, _, \mathsf{ret}(v))$*.*
  - *Each vertex that reads* $x$ *has at most one corresponding read dependency:* $\forall \nu, \nu', \nu'', x.\, \nu \xrightarrow{\mathsf{WR}_x} \nu' \wedge \nu'' \xrightarrow{\mathsf{WR}_x} \nu \implies \nu = \nu''$*.*
  - *Each vertex that is read from is visible:* $\forall \nu, x.\, \nu \xrightarrow{\mathsf{WR}_x} _ \implies \mathsf{vis}(\nu)$*.*

  *Informally,* $\nu \xrightarrow{\mathsf{WR}_x} \nu'$ *means that* $\nu'$ *reads what* $\nu$ *wrote to* $x$*.*
- $\mathsf{WW} : \mathsf{Reg} \to 2^{\mathcal{V} \times \mathcal{V}}$ *specifies per-register* write-dependency *relations, such that for each* $x \in \mathsf{Reg}$*,* $\mathsf{WW}_x$ *is an irreflexive total order on* $\{\nu \in \mathcal{V} \mid \mathsf{vis}(\nu) \wedge (_, _, \mathsf{write}(x, _)) \in \nu\}$*. Informally,* $\nu \xrightarrow{\mathsf{WW}_x} \nu'$ *means that* $\nu'$ *overwrites what* $\nu$ *wrote to* $x$*.*
- $\mathsf{RW} \in \mathsf{Reg} \to 2^{\mathcal{V} \times \mathcal{V}}$ *specifies per-register* anti-dependency *relations:*

$$\nu \xrightarrow{\mathsf{RW}_x} \nu' \iff \nu \neq \nu' \wedge ((\exists \nu''.\, \nu'' \xrightarrow{\mathsf{WW}_x} \nu' \wedge \nu'' \xrightarrow{\mathsf{WR}_x} \nu) \vee (\mathsf{vis}(\nu') \wedge (_, _, \mathsf{write}(x, _)) \in \nu' \wedge (_, _, \mathsf{ret}(x, v_{\mathsf{init}})) \in \nu)).$$

  *Informally,* $\nu \xrightarrow{\mathsf{RW}_x} \nu'$ *means that* $\nu'$ *overwrites the write to* $x$ *that* $\nu$ *previously read (the initial value of* $x$ *is overwritten by any write to* $x$*).*
- $\mathsf{PO}, \mathsf{CL} \in 2^{\mathcal{V} \times \mathcal{V}}$ *are the per-thread and client orders lifted to pairs of graph vertexes: e.g.,* $\nu \xrightarrow{\mathsf{PO}(H)} \nu' \iff \exists \alpha \in \nu, \alpha' \in \nu'.\, \alpha <_{\mathsf{po}(H)} \alpha'$*.*

We let $\mathsf{Graph}(H)$ denote the set of all opacity graphs of $H$. We say that a graph $G$ is *acyclic*, written $\mathsf{acyclic}(G)$, if its edges do not form a directed cycle. We also refer to histories resulting from topological sortings of vertexes in a graph $G$ as its *linearizations* and denote their set by $\mathsf{lins}(G)$. The next lemma shows that we can check privatization-safe opacity of a history by checking its consistency and the acyclicity of its opacity graph, with any linearization of the graph yielding a matching atomic history.

▶ **Lemma 13.** $\forall H.\, (\mathsf{cons}(H) \wedge \exists G \in \mathsf{Graph}(H).\, \mathsf{acyclic}(G)) \implies \mathsf{lins}(G) \subseteq \mathcal{H}_{\mathsf{atomic}}$.

The lemma is proven analogously to Lemma 6.4 in [26, §B.2]. It implies the following theorem, which gives a criterion for the privatization-safe opacity of a TM $\mathcal{H}$.

▶ **Theorem 14.** *$\mathcal{H} \sqsubseteq \mathcal{H}_{\mathsf{atomic}}$ holds if $\forall H \in \mathcal{H}.\, \mathsf{cons}(H) \wedge \exists G \in \mathsf{Graph}(H).\, \mathsf{acyclic}(G)$.*

In comparison to the graph characterization of the ordinary opacity [21], ours is more complex: the graph includes non-transactional accesses and the acyclicity check has to take into account paths involving them. We now formulate lemmas that simplify reasoning about non-transactional operations: they allow proving the privatization-safe opacity of a TM using Theorem 14 with only small adjustments to a proof of its ordinary opacity using graph characterization. The latter characterization includes only transactions as nodes of the graph, but additionally considers paths including the lifting of the real-time order from §4 to transactions: for a history $H$, we let $\mathsf{RT}(H)$ be the relation between transactions in $H$ such that $T <_{\mathsf{RT}(H)} T'$ iff for some $\alpha \in T$ and $\alpha' \in T$ we have $\alpha <_{\mathsf{rt}(H)} \alpha'$. We also let $\mathsf{DEP}$ denote any edge in a given graph $G$, and we let $\mathsf{txDEP}$ denote an edge between two transactions.

The following lemma exploits CDRF to show that, for every path between two transactions in an acyclic opacity graph, there is another path replacing edges involving non-transactional accesses by real-time order edges or transactional dependencies.

▶ **Lemma 15.** *Consider an acyclic opacity graph $G = (\mathcal{V}, \mathsf{vis}, \mathsf{WR}, \mathsf{WW}, \mathsf{RW}, \mathsf{PO}, \mathsf{CL})$ of a consistent CDRF history $H$. For any two transactions $T$ and $T'$, if $T \xrightarrow{\mathsf{DEP}}{}^{*} T'$, then $T \xrightarrow{\mathsf{RT} \cup \mathsf{txDEP}}{}^{*} T'$.*

The next lemma exploits CDRF to show that, for every path between a transaction and a non-transactional access in an acyclic opacity graph, there is another path where per-thread order is the only kind of an edge between transactions and non-transactional accesses.

▶ **Lemma 16.** *Consider an acyclic opacity graph $G = (\mathcal{V}, \mathsf{vis}, \mathsf{WR}, \mathsf{WW}, \mathsf{RW}, \mathsf{PO}, \mathsf{CL})$ of a CDRF history $H$. For any transaction $T$ and non-transactional access $n$:*

- *if $T \xrightarrow{\mathsf{DEP}}{}^{*} n$, then there are $T'$ and $n'$ such that $T \xrightarrow{\mathsf{RT} \cup \mathsf{txDEP}}{}^{*} T' \xrightarrow{\mathsf{PO}} n' \xrightarrow{\mathsf{CL}}{}^{*} n$;*
- *if $n \xrightarrow{\mathsf{DEP}}{}^{*} T$, then there are $T'$ and $n'$ such that $n \xrightarrow{\mathsf{CL}}{}^{*} n' \xrightarrow{\mathsf{PO}} T' \xrightarrow{\mathsf{RT} \cup \mathsf{txDEP}}{}^{*} T$.*

Our method for proving the privatization-safe opacity of a TM (which we illustrate in §7) uses Lemmas 15 and 16 to reduce proving the acyclicity of an opacity graph to proving the absence of cycles in the projection of the graph to transactions, enriched with real-time order edges. The simplified acyclicity check is exactly the one required in the graph characterization of the ordinary opacity [21], allowing us to reuse existing proofs.

In the following we prove Lemmas 15 and 16. We show the existence of the paths required in the lemmas by using CDRF to eliminate $\mathsf{WR}/\mathsf{WW}/\mathsf{RW}$-dependencies between transactions and non-transactional accesses. Each of the dependencies to be eliminated corresponds to a conflict in a matching atomic history, which CDRF guarantees to relate by happens-before. The next lemma exploits this observation.

▶ **Lemma 17.** *Consider an acyclic opacity graph* $G = (\mathcal{V}, \mathsf{vis}, \mathsf{WR}, \mathsf{WW}, \mathsf{RW}, \mathsf{PO}, \mathsf{CL})$ *of a consistent CDRF history* $H$*. For any transaction* $T$ *and non-transactional access* $n$*:*
1. *if* $T \xrightarrow{\mathsf{DEP}} n$*, then there are* $T'$ *and* $n'$ *such that* $T \xrightarrow{\mathsf{DEP}}{}^{*} T' \xrightarrow{\mathsf{PO}} n' \xrightarrow{\mathsf{CL}}{}^{*} n$*;*
2. *if* $n \xrightarrow{\mathsf{DEP}} T$*, then there are* $T'$ *and* $n'$ *such that* $n \xrightarrow{\mathsf{CL}}{}^{*} n' \xrightarrow{\mathsf{PO}} T' \xrightarrow{\mathsf{DEP}}{}^{*} T$*.*

For example, consider an execution of the program in Figure 1 where $T_2$ reads `false` from `priv` and writes to `x` before $n$ does. The corresponding acyclic graph contains both $T_2 \xrightarrow{\mathsf{WW}_{\mathtt{x}}} n$ and $T_2 \xrightarrow{\mathsf{RW}_{\mathtt{priv}}} T_1 \xrightarrow{\mathsf{PO}} n$. To prove Lemma 17, we lift $<_{\mathsf{po}(H)}$, $<_{\mathsf{ef}(H)}$, $<_{\mathsf{cl}(H)}$ and $<_{\mathsf{hb}(H)}$ from Definition 2 to vertexes of the graph as expected, writing $<_{\mathsf{PO}(H)}$, $<_{\mathsf{EF}(H)}$, $<_{\mathsf{CL}(H)}$ and $<_{\mathsf{HB}(H)}$ for the resulting relations. We also write $\leq$ for their reflexive closure. We rely on the following easy result (proved in §D).

▶ **Proposition 18.** *In a TDRF history* $H$*, for any* $T$ *and* $n$ *we have:*
- *if* $T <_{\mathsf{HB}(H)} n$*, then there are* $T'$ *and* $n'$ *such that* $T \leq_{\mathsf{EF}(H)} T' <_{\mathsf{PO}(H)} n' \leq_{\mathsf{CL}(H)} n$*;*
- *if* $n <_{\mathsf{HB}(H)} T$*, then there are* $T'$ *and* $n'$ *such that* $n \leq_{\mathsf{CL}(H)} n' <_{\mathsf{PO}(H)} T' \leq_{\mathsf{EF}(H)} T$*.*

**Proof of Lemma 17.** We only prove part 1, as part 2 can be proven analogously. Assume $T \xrightarrow{\mathsf{DEP}} n$. If $T \xrightarrow{\mathsf{PO}} n$, then $T \xrightarrow{\mathsf{DEP}}{}^{*} T \xrightarrow{\mathsf{PO}} n \xrightarrow{\mathsf{CL}}{}^{*} n$, which trivially concludes the proof. In the following, we consider the remaining case when $\neg(T \xrightarrow{\mathsf{PO}} n)$ and $T \xrightarrow{\mathsf{WR} \cup \mathsf{RW} \cup \mathsf{WW}} n$, so that $T$ and $n$ contain conflicting actions. Let $\mathcal{A}$ denote the following set of pairs $(T', n')$ of a transaction and a non-transactional access:

$$\mathcal{A} \triangleq \{(T', n') \mid \exists L \in \mathsf{lins}(G).\, T \leq_{\mathsf{EF}(L)} T' <_{\mathsf{PO}(L)} n' \leq_{\mathsf{CL}(L)} n\}.$$

By Definition 12, for any $(T', n') \in \mathcal{A}$ we have $T' \xrightarrow{\mathsf{PO}} n' \xrightarrow{\mathsf{CL}}{}^{*} n$. It suffices to show that there is $T'$ such that $T \xrightarrow{\mathsf{DEP}}{}^{*} T'$ and $(T', _) \in \mathcal{A}$. Proceeding by contradiction, let us assume that this is not the case: for every $(T', _) \in \mathcal{A}$, there is no edge $T \xrightarrow{\mathsf{DEP}}{}^{*} T'$ in $G$. Then extending the graph with edges $\{T' \xrightarrow{\mathsf{DEP}} T \mid (T', _) \in \mathcal{A}\}$ will not introduce a cycle. Hence, there exists a linearization $L \in \mathsf{lins}(G)$ in which every $(T', _) \in \mathcal{A}$ occurs before $T$: $\forall T'.\, (T', _) \in \mathcal{A} \implies T' <_{\mathsf{EF}(L)} T$.

Since, the history $H$ is consistent and has an acyclic opacity graph $G$, by Lemma 13 we get $L \in \mathsf{lins}(G) \subseteq \mathcal{H}_{\mathsf{atomic}}$. Since $H$ is CDRF, the conflicting pair $T$ and $n$ are ordered by $\mathsf{HB}(L)$. Moreover, since $T$ occurs before $n$ in $L$ and $\mathsf{HB}(L)$ is consistent with the execution order of $L$, we have $T <_{\mathsf{HB}(L)} n$. From this by Proposition 18, for some $T''$ and $n''$ we have $T \leq_{\mathsf{EF}(L)} T'' <_{\mathsf{PO}(L)} n'' \leq_{\mathsf{CL}(L)} n$. Hence, $T \leq_{\mathsf{EF}(L)} T''$ and $(T'', n'') \in \mathcal{A}$. But by the construction of $L$ we have $T'' <_{\mathsf{EF}(L)} T$, which contradicts the definition of $\mathsf{ef}$ as a total order on transactions. This contradiction demonstrates the required. ◀

The following result leverages Lemma 17 to show that, for every path between two transactions in an acyclic opacity graph, there is another path replacing some edges involving non-transactional accesses by real-time order edges or transactional dependencies.

▶ **Lemma 19.** *Consider an acyclic opacity graph* $G = (\mathcal{V}, \mathsf{vis}, \mathsf{WR}, \mathsf{WW}, \mathsf{RW}, \mathsf{PO}, \mathsf{CL})$ *of a consistent CDRF history* $H$*. For any two transactions* $T$ *and* $T'$*, if* $T \xrightarrow{\mathsf{DEP}}{}^{+} T'$*, then there are two transactions* $T_1$ *and* $T_2$ *such that* $T \xrightarrow{\mathsf{DEP}}{}^{*} T_1 \xrightarrow{\mathsf{txDEP} \cup \mathsf{RT}} T_2 \xrightarrow{\mathsf{DEP}}{}^{*} T'$*.*

**Proof.** Assume $T \xrightarrow{\mathsf{DEP}}{}^{+} T'$ and consider the corresponding path in the graph $G$. If there are no non-transactional accesses on this path, then $T \xrightarrow{\mathsf{txDEP}}{}^{+} T'$, so the lemma holds trivially.

Assume now that there are non-transactional accesses on the path corresponding to $T \xrightarrow{\mathsf{DEP}}{}^{+} T'$. Let $n$ and $n'$ be the first and the last such accesses respectively, and also let $T'_1$ ($T'_2$) be the transaction immediately preceding $n$ (following $n'$) on the path. Since $G$ is acyclic and CL relates every pair of non-transactional accesses, we must have $n \xrightarrow{\mathsf{CL}}{}^* n'$. Then $T \xrightarrow{\mathsf{DEP}}{}^* T'_1 \xrightarrow{\mathsf{DEP}} n \xrightarrow{\mathsf{CL}}{}^* n' \xrightarrow{\mathsf{DEP}} T'_2 \xrightarrow{\mathsf{DEP}}{}^* T'$. Applying Lemma 17(1) to $T'_1 \xrightarrow{\mathsf{DEP}} n$ and Lemma 17(2) to $n' \xrightarrow{\mathsf{DEP}} T'_2$, we get that there are $T_1$, $n_1$, $T_2$ and $n_2$ such that:

$$T \xrightarrow{\mathsf{DEP}}{}^* T'_1 \xrightarrow{\mathsf{DEP}}{}^* T_1 \xrightarrow{\mathsf{PO}} n_1 \xrightarrow{\mathsf{CL}}{}^* n \xrightarrow{\mathsf{CL}}{}^* n' \xrightarrow{\mathsf{CL}} n_2 \xrightarrow{\mathsf{PO}} T_2 \xrightarrow{\mathsf{DEP}}{}^* T'_2 \xrightarrow{\mathsf{DEP}}{}^* T'.$$

Then $T \xrightarrow{\mathsf{DEP}}{}^* T_1 \xrightarrow{\mathsf{PO}} n_1 \xrightarrow{\mathsf{CL}}{}^* n_2 \xrightarrow{\mathsf{PO}} T_2 \xrightarrow{\mathsf{DEP}}{}^* T'$. By Definition 12 of PO and CL, $T_1$ ends before $T_2$ starts, so that $T_1 \xrightarrow{\mathsf{RT}} T_2$. Then $T \xrightarrow{\mathsf{DEP}}{}^* T_1 \xrightarrow{\mathsf{RT}} T_2 \xrightarrow{\mathsf{DEP}}{}^* T'$, as required. ◀

**Proof of Lemma 15.** To prove the lemma, we iteratively construct a path in $G$ demonstrating that $T \xrightarrow{\mathsf{RT}\cup\mathsf{txDEP}}{}^* T'$. At the $k$-th iteration we construct a sequence $\pi_k$ of transactions $T_0, T'_0, T_1, T'_1, \dots, T_k, T'_k \in \mathcal{V}$ such that:

- $T_0 = T$, $T'_k = T'$, and
- $T_0 \xrightarrow{\mathsf{DEP}}{}^* T'_0 \xrightarrow{\mathsf{RT}\cup\mathsf{txDEP}} T_1 \xrightarrow{\mathsf{DEP}}{}^* T'_1 \xrightarrow{\mathsf{RT}\cup\mathsf{txDEP}} \dots \xrightarrow{\mathsf{RT}\cup\mathsf{txDEP}} T_k \xrightarrow{\mathsf{DEP}}{}^* T'_k$.

We start the construction with a sequence $\pi_0 = T, T'$, which satisfies the above conditions because $T \xrightarrow{\mathsf{DEP}}{}^* T'$. We stop the construction once we get a sequence $\pi_k$ such that $T_i = T'_i$ for each $i = 0..k$: in this case the sequence yields a path of the required form. Otherwise, we construct $\pi_{k+1}$ from $\pi_k$ as follows. We choose any two transactions $T_i$ and $T'_i$ in $\pi_k$ such that $T_i \neq T'_i$ and, hence, $T_i \xrightarrow{\mathsf{DEP}}{}^+ T'_i$. By Lemma 19, there are $T''_i$ and $T'''_i$ such that $T_i \xrightarrow{\mathsf{DEP}}{}^* T''_i \xrightarrow{\mathsf{txDEP}\cup\mathsf{RT}} T'''_i \xrightarrow{\mathsf{DEP}}{}^* T'_i$. Then we let $\pi_{k+1} = T_0, T'_0, \dots, T_i, T''_i, T'''_i, T'_i, \dots, T_k, T'_k$.

Since $G$ is acyclic, in any $\pi_k$ the only transactions that can coincide are some consecutive $T_i$ and $T'_i$. Thus, $\pi_k$ contains at least $k+1$ distinct transactions. But then our transformation has to stop after at most $n$ steps, where $n$ is the number of transactions in $G$. ◀

**Proof of Lemma 16.** We only prove part 1, as part 2 can be proven analogously. Assume $T \xrightarrow{\mathsf{DEP}}{}^* n$. Then there are $T''$ and $n''$ such that $T \xrightarrow{\mathsf{DEP}}{}^* T'' \xrightarrow{\mathsf{DEP}} n'' \xrightarrow{\mathsf{DEP}}{}^* n$. By Lemma 17, there are $T'$ and $n'$ such that $T \xrightarrow{\mathsf{DEP}}{}^* T'' \xrightarrow{\mathsf{DEP}}{}^* T' \xrightarrow{\mathsf{PO}} n' \xrightarrow{\mathsf{CL}}{}^* n'' \xrightarrow{\mathsf{CL}}{}^* n$. Then $T \xrightarrow{\mathsf{DEP}}{}^* T' \xrightarrow{\mathsf{PO}} n' \xrightarrow{\mathsf{CL}}{}^* n$. By Lemma 15, $T \xrightarrow{\mathsf{RT}\cup\mathsf{txDEP}}{}^* T'$, implying the required. ◀

As we show in §D, the observations in the proofs of the Lemmas 15 and 16 additionally let us establish the following interesting theorem, giving an equivalent formulation of CDRF in terms of dependencies between transactions.

▶ **Theorem 20.** *Given a consistent history $H$, $\mathsf{CDRF}(H)$ holds if and only if in each acyclic opacity graph $G = (\mathcal{V}, \mathsf{vis}, \mathsf{WR}, \mathsf{WW}, \mathsf{RW}, \mathsf{PO}, \mathsf{CL}) \in \mathsf{Graph}(H)$ there is a path over edges from $\mathsf{PO} \cup \mathsf{CL} \cup \mathsf{txDEP} \cup \mathsf{RT}(H)$ between every pair of vertexes containing conflicting actions.*

## 7 Case Study: FencedTL2

In this section we illustrate how Lemmas 15 and 16 enable simple proofs of privatization-safe opacity using an example of a privatization-safe version of TL2 [14]. We give only the key parts of the proof and defer details to §E. There we also give a proof of privatization-safe opacity of a TM based on two-phase locking [21], which is privatization-safe.

As we noted in §1, the TL2 algorithm by itself is not privatization-safe. The reason is that TL2 executes transactions optimistically, buffering their writes, and flushes them to memory only on commit. Thus, in the example in Figure 1, it is possible for the transaction $T_1$ to privatize x and for $n$ to modify it after $T_2$ started committing, but before its write to x reached the memory, so that $T_2$'s write subsequently overwrites $n$'s write and violates the postcondition. We can make TL2 privatization-safe by modifying its implementation so that it executes a *transactional fence* [30, 34] at the end of every transaction, an implementation we call *FencedTL2*. The fence has a semantics similar to *Read-Copy-Update (RCU)* [31]: it blocks until all the concurrent transactions that were active when the fence was invoked complete, by either committing or aborting. For instance, in the example in Figure 1 executing a transactional fence after $T_1$ would block the thread until $T_2$ commits or aborts, thus ensuring that $n$ is not overwritten by $T_2$'s buffered write. The above way of making a TM privatization-safe is used in the GCC compiler [18] (albeit with TinySTM [16] instead of TL2) and has been experimentally evaluated in [35, 36].

To prove privatization-safe opacity of FencedTL2, for every one of its executions we inductively construct an opacity graph (with added real-time order edges) that matches its history. This is done with the help of the following *graph updates*, which specify how and when in the execution to extend the graph:

- At the start of a transaction $T$, a graph update $\textsc{txinit}(T)$ adds a new vertex $T$ and extends the real-time order with edges $T' \xrightarrow{\mathsf{RT}} T$ for every completed transaction $T'$.
- At the end of a read operation of a transaction $T$ reading from an object $x$, a graph update $\textsc{txread}(T, x)$ adds a read dependency $\nu \xrightarrow{\mathsf{WR}_x} T$, where $\nu$ is the vertex that wrote the value returned by the read.
- During the commit of a transaction $T$, TL2 validates the consistency of $T$'s read-set before flushing $T$'s write-set into memory. At the last step of the validation, a graph update $\textsc{txwrite}(T, x)$ adds a write dependency $\nu \xrightarrow{\mathsf{WW}_x} T$ for every object $x$ in the write-set of $T$, where $\nu$ is the vertex that wrote the previous value of $x$.
- Upon each non-transactional write $n$ to an object $x$, a graph update $\textsc{ntxwrite}(n, x)$ adds a new vertex $n$ and a write-dependency $\nu \xrightarrow{\mathsf{WW}_x} n$, where $\nu$ is the vertex that wrote the previous value of $x$.
- Upon each non-transactional read $n$ from an object $x$, a graph update $\textsc{ntxread}(n, x)$ adds a new vertex $n$ and a read dependency $\nu \xrightarrow{\mathsf{WR}_x} n$, where $\nu$ is the vertex that wrote the value returned by $n$.

The updates also add anti-dependencies of the form $_ \xrightarrow{\mathsf{RW}} T$ induced by new read- and write-dependencies.

At each step of the graph construction we prove that the graph remains acyclic. Then Theorem 14 guarantees that the history of the execution is opaque. We use Lemmas 15 and 16 to reduce the task of proving the graph acyclicity to proving the absence of cycles involving transactions only. To discharge the latter proof obligation, we reuse our previous proof of opacity of TL2 [26], also done via the graph characterization. This proof establishes the following invariant over pairs $(H, G)$ of a history $H$ and a graph $G$:

- $\mathsf{INV}_1$: $H$ is a consistent history and the relation $\mathsf{txDEP} \cup \mathsf{RT}$ is acyclic.

To enable the reduction from privatization-safe to ordinary opacity, we prove the following invariant, which states the guarantee provided by fences in FencedTL2:

- $\mathsf{INV}_2$: For every uncompleted transaction $T$ and a transaction $T'$, $T \xrightarrow{\mathsf{txDEP}}{}^* T' \xrightarrow{\mathsf{PO}} _$ does not hold.

An informal justification of $\mathsf{INV}_2$ is as follows. By construction of the graph it is possible to establish that $T'$ can depend on an uncompleted transaction $T$ only when they execute concurrently. In this case, the fence of $T'$ will wait for $T$ to commit or abort, and until then there cannot be any transactions or non-transactional accesses in the thread of $T'$ later in the per-thread order. By Theorem 14, privatization-safe opacity of FencedTL2 follows from

▶ **Theorem 21.** $\forall H \in \text{FencedTL2}.\, \mathsf{CDRF}(H) \implies \exists G.\, (H, G) \in \mathsf{INV}_1 \wedge \mathsf{INV}_2 \wedge \mathsf{acyclic}(G)$.

We prove Theorem 21 by induction on the length of the TM execution inducing $H$, constructing $G$ as described above and showing that it remains acyclic after each update with the aid of the two invariants. Due to space constraints, we only explain how we prove acyclicity in the case of a graph update TXWRITE, which illustrates the use of Lemmas 15 and 16.

▶ **Lemma 22.** *Let $(H', G')$ be the result of performing an update* TXWRITE$(T, x)$ *on $(H, G)$. Assume that $(H, G), (H', G') \in \mathsf{INV}_1 \wedge \mathsf{INV}_2$ and $G$ is acyclic. Then $G'$ is acyclic too.*

**Proof.** By contrapositive: we assume that $G'$ contains a simple cycle and show that $G'$ violates either $\mathsf{INV}_1$ or $\mathsf{INV}_2$. The graph update adds an edge of the form $_ \xrightarrow{\mathsf{WW}_x} T$ and the derived edges of the form $_ \xrightarrow{\mathsf{RW}_x} T$. Since both kinds of edges end in the same vertex $T$, they cannot occur in the same simple cycle. Hence, we can consider them separately.

Consider a simple cycle involving a new edge $\nu \xrightarrow{\mathsf{DEP}} T$ for some vertex $\nu$. By our assumption, there must be a reverse path $T \xrightarrow{\mathsf{DEP}}{}^* \nu$ in $G$. Let us first consider the case when $\nu$ is a transaction $T'$. Since $G$ is acyclic and $H$ is consistent and CDRF, by Lemma 15 the path $T \xrightarrow{\mathsf{DEP}}{}^* T'$ can be reduced to $T \xrightarrow{\mathsf{RT} \cup \mathsf{txDEP}}{}^* T'$. Since $G'$ only extends $G$, the same path is present in $G'$ too. Then $T' \xrightarrow{\mathsf{txDEP}} T \xrightarrow{\mathsf{RT} \cup \mathsf{txDEP}}{}^* T'$ is a cycle over transactions in $G'$, which contradicts $(H', G') \in \mathsf{INV}_1$. We now consider the case when $\nu$ is a non-transactional access $n$. Since $G$ is acyclic and $H$ is consistent and CDRF, by Lemma 16 there exist $T'$ and $n'$ such that $T \xrightarrow{\mathsf{txDEP}}{}^* T' \xrightarrow{\mathsf{PO}} n' \xrightarrow{\mathsf{CL}}{}^* n$ holds in $G$. Note that $T$ is an uncompleted transaction, since it currently performs a graph update. Therefore, $T \xrightarrow{\mathsf{txDEP}}{}^* T' \xrightarrow{\mathsf{PO}} n' \xrightarrow{\mathsf{CL}}{}^* n$ is a contradiction to $(H, G) \in \mathsf{INV}_2$. ◀

## 8 The Cost of Privatization-Safety

We now present a result about the inherent cost of privatization-safety, by which we mean guaranteeing strongly atomic semantics to TDRF programs. In addition to TM histories, we consider the prefix-closed set of all TM *executions* $\mathcal{X}$, ranged over by $\varphi$. Unlike histories, they include internal TM actions that only occur in transactions and are not a part of the TM interface. One type of an internal action are *write-backs* of the form $(a, t, \mathsf{wb}(x, v))$, where $a \in \mathsf{ActionId}$, $t \in \mathsf{ThreadID}$, $x \in \mathsf{Reg}$, and $v \in \mathbb{Z}$. A write-back denotes a transaction of a thread $t$ writing a value $v$ to a register $x$. We assume that a TM implementation is represented by a pair $(\mathcal{H}, \mathcal{X})$ of a set of histories and a set of executions producing them.

▶ **Definition 23.** *A TM system $(\mathcal{H}, \mathcal{X})$ is* progressive *when for any $\varphi \in \mathcal{X}$ with at most one uncompleted transaction $T$, if the last interface action by $T$ in $\varphi$ is a request $\alpha$, there exists a sequence of internal TM actions $\varphi'$ by $T$ and a response $\alpha'$ matching $\alpha$ such that $\varphi\varphi'\alpha' \in \mathcal{X}$.*

▶ **Definition 24.** *A TM system $(\mathcal{H}, \mathcal{X})$ has* invisible reads *when for any $\varphi\varphi' \in \mathcal{X}$ such that $\varphi$ contains at most one uncompleted transaction $T$ and $\varphi'$ is a sequence of actions corresponding to another uncompleted transaction $T'$ only conflicting with reads by $T$, if the last interface action by $T'$ is a request $\alpha$, there exists a sequence of internal TM actions $\varphi''$ by $T'$ and a response $\alpha' \neq (_, _, \mathsf{aborted})$ matching $\alpha$ such that $\varphi\varphi'\varphi''\alpha' \in \mathcal{X}$.*

Our progressiveness property is analogous to obstruction-freedom [22], requiring a transaction to complete when running solo. Our invisible reads property can be ensured when the TM only writes to thread-local memory upon reading [21]. The FencedTL2 TM from §7 is privatization-safe and has invisible reads, but is not progressive due to its use of fences. As the following theorem shows, this is not accidental.

▶ **Theorem 25.** *A TM system that guarantees strongly atomic semantics to TDRF programs cannot both be progressive and have invisible reads.*

We rely on the following proposition, proved in §F.

▶ **Proposition 26.** *Consider a TM system that guarantees strongly atomic semantics to TDRF programs. If $\varphi$ is a TM execution of a single atomic block where the latter commits, and $(_, _, \mathsf{write}(x, v))$ is its last write request to $x$, then $\varphi$ also contains a write-back $(_, _, \mathsf{wb}(x, v))$, and all write-backs to $x$ occur in $\varphi$ after the first write request to $x$.*

**Proof of Theorem 25.** The proof is by contradiction. Assume there exists a progressive TM $(\mathcal{H}, \mathcal{X})$ with invisible reads that guarantees strong atomicity to every TDRF program $P$, so that $[\![P]\!](\mathcal{H}) \preceq [\![P]\!](\mathcal{H}_{\mathsf{atomic}})$. We choose a particular TDRF program $P$ and construct a counterexample trace from $[\![P]\!](\mathcal{H})$ that does not have a matching trace in $[\![P]\!](\mathcal{H}_{\mathsf{atomic}})$. Namely, we consider the following program $P$, similar to the one in Figure 1:

$$\{\, \mathtt{priv} = \mathtt{false} \wedge \mathtt{x} = 0 \,\}$$

| | |
|---|---|
| `l`$_1$ `= atomic {` | `atomic {` |
| `  priv = true; }` $//T_1$ | `  if (!priv)` |
| `if (l`$_1$ `== committed)` | `    x = 42;` |
| `  l`$_2$ `= x;` $//n$ | `}` $//T_2$ |

We first consider a single-threaded program executing the atomic block in the right-hand-side thread $t_2$ of $P$. The TM always allows the program to execute requests (§2), and the invisible reads property ensures that the TM responds to them without aborting. Therefore, there is an execution $\varphi_2^0 \in \mathcal{X}$ consisting only of actions of the atomic block of $t_2$ in $P$ ending with a commit-response. By Proposition 26, the execution of $\varphi_2^0$ contains a write-back $(_, t_2, \mathsf{wb}(\mathtt{x}, 42))$. Let $\varphi_2$ be the prefix of $\varphi_2^0$ until the first write-back $w = (_, t_2, \mathsf{wb}(\mathtt{x}, 42))$. By Proposition 26, $\varphi_2$ contains a write request to `x` and, therefore, a preceding response $(_, t_2, \mathsf{ret}(\mathtt{false}))$ to a read from `priv`. The set of TM executions is prefix-closed, so $\varphi_2 w \in \mathcal{X}$.

Note that $\varphi_2$ corresponds to a (partial) trace of $P$. We now let $P$ continue $\varphi_2$ by executing the atomic block of the left-hand-side thread $t_1$. The TM always allows $t_1$ to execute requests (§2), and the invisible reads property ensures that the TM responds to them without aborting, as they only conflict with $t_2$'s read from `priv` in $\varphi_2$. We thus obtain a sequence of actions $\varphi_1$ corresponding to a committed transaction $T_1$ such that $\varphi_2\varphi_1 \in \mathcal{X}$. We can then execute $n = (_, t_1, \mathsf{read}(\mathtt{x}))(_, t_1, \mathsf{ret}(0))$, which returns the initial value of `x` as there has not been any write-back to `x` yet. We thereby obtain an execution $\varphi_2\varphi_1 n \in \mathcal{X}$ in which thread $t_1$ of $P$ has executed to completion.

We now let $P$ resume executing the atomic block of thread $t_2$. Since the TM is progressive, the execution $\varphi_2\varphi_1 n$ can be extended to an execution $\varphi = \varphi_2\varphi_1 n \varphi_2' \in \mathcal{X}$ where the atomic block is completed, yielding a transaction $T_2$. We first consider the case when $T_2$ commits in $\varphi$. The execution $\varphi$ corresponds to a trace $\tau \in [\![P]\!](\mathcal{H})$. Since $[\![P]\!](\mathcal{H}) \preceq [\![P]\!](\mathcal{H}_{\mathsf{atomic}})$, there exists a trace $\tau' \in [\![P]\!](\mathcal{H}_{\mathsf{atomic}})$ matching $\tau$. Above we established that $\varphi_2$ reads `false` from `priv` and, hence, so does $T_2$. To justify reading this value in $\tau'$, $T_2$ must commit in this trace before $T_1$ starts and, therefore, before $n$ starts too. Hence, $n$ must observe $T_2$'s write to `x` in $\tau'$, even though it observes the initial value in $\tau$. Then $\tau'$ cannot match $\tau$, and this contradiction concludes the proof.

We now consider the case when $T_2$ aborts in $\varphi$. Above we established that $\varphi_2^0 = \varphi_2 w _ \in \mathcal{X}$, so that $\varphi_2 w \in \mathcal{X}$. Since the TM executes write-backs as atomic writes, if a transaction is interrupted when a write-back $w$ is pending, it proceeds with $w$ once its execution resumes. Hence, it must be the case that $\varphi_2'$ takes the form of $w\varphi_2''$, so that $\varphi = \varphi_2\varphi_1 n w \varphi_2''$. Since the TM does not impose restrictions on the placement of the non-transactional accesses (§2), it must also allow an execution $\varphi_2 \varphi_1 w n' \varphi_2'' \in \mathcal{X}$, where $n' = (_, t_1, \mathsf{read}(\mathtt{x}))(_, t_1, \mathsf{ret}(42))$ returns the value written by $w$. This execution corresponds to a trace $\tau \in [\![P]\!](\mathcal{H})$. Since $[\![P]\!](\mathcal{H}) \preceq [\![P]\!](\mathcal{H}_{\mathsf{atomic}})$, there exists a trace $\tau' \in [\![P]\!](\mathcal{H}_{\mathsf{atomic}})$ matching $\tau$. In this trace $n'$ reads 42 written by an aborted transaction $T_2$, which cannot happen under $\mathcal{H}_{\mathsf{atomic}}$. Hence, $\tau' \notin [\![P]\!](\mathcal{H}_{\mathsf{atomic}})$, and this contradiction concludes the proof. ◀

## 9 Related Work and Discussion

We have previously proposed a notion of DRF for privatization-unsafe TMs and a corresponding variant of opacity that ensure the Fundamental Property [26]. This work considered a more low-level programming model, which required inserting fences after some of the transactions for a program to be DRF. The resulting DRF notion was thus more involved than TDRF. Showing that the simpler TDRF is enough for privatization-safe TMs required us to address new technical challenges, such as the need to generalize TDRF to concurrent histories (to formulate privatization-safe opacity, §4) and to prove the delicate path reduction lemmas linking TDRF with properties of opacity graphs (§6). Furthermore, unlike [26], our results are also applicable to TMs that achieve privatization-safety by means other than fences, such as a lock-based TM we handle in §E.2. Our results also suggest a strengthening of those in [26]; we defer the details to §G.

The notion of TDRF we use is a variant of the one proposed by Dalessandro et al. [11]. They also suggested that the notion should satisfy the Fundamental Property, but with strict serializability as the required condition on the TM. As we argued in §4, this condition is too strong, as it does not allow the proofs of TM correctness to benefit from the DRF of programs using it. In this paper we justify the appropriateness of TDRF by proposing a matching TM correctness condition that enables proofs of common TMs and proving the Fundamental Property for it. This also requires us to generalize TDRF to concurrent histories.

In this paper we assumed sequential consistency as a baseline non-transactional memory model. However, transactions are being integrated into languages, such as C++, that have weaker memory models [1]. Transactional sequential consistency, which we use as our strongly atomic semantics, is equivalent to that prescribed by the C++ memory model without relaxed transactions or non-SC atomics [9], and our definition of a data race is given in the axiomatic style used in the C++ memory model [2]. Hence, we believe that in the future our results can be generalized to the wider C++ model, in particular, by weakening the client order in Definition 2 to account for non-SC non-transactional accesses.

Abadi et al. also proposed disciplines for privatization with a formal justification of their safety [3, 4]. However, these disciplines are more restrictive than ours: they either prohibit mixing transactional and non-transactional accesses to the same register [4] or require explicit commands to privatize and publish an object [3]. Such disciplines are particular ways of achieving the more general notion of TDRF that we adopted.

Attiya and Hillel [7] investigated the cost of privatization in progressive TMs. Unlike us, they considered support for privatization to be part of TM interface and did not rely on a formal notion of privatization-safety. They proved the impossibility of supporting privatization in eager TMs, and a lower bound on its implementation cost in lazy TMs. Our Theorem 25 unifies and strengthens their results, as it states the impossibility of providing privatization-safety for all progressive TMs with invisible reads. We also make the results more rigorous by linking them to a formal notion of privatization-safety based on TDRF.

## Table of Annexes

# A Additional Details for §2: Programming Language Semantics

## A.1 Formal Definition of Well-formed Traces

To formalize restrictions on accesses to variables by primitive commands, we partition the set $\mathsf{PCom}$ into $m$ classes: $\mathsf{PCom} = \biguplus_{t=1}^{m} \mathsf{LPcomm}_t$. The intention is that commands from $\mathsf{LPcomm}_t$ can access only the local variables of thread $t$ ($\mathsf{LVar}_t$). To ensure that in our programming language a thread $t$ does not access local variables of other threads, we require that the thread cannot mention such variables in the conditions of `if` and `while` commands and can only use primitive commands from $\mathsf{LPcomm}_t$.

▶ **Definition 27.** *A trace $\tau$ is a finite sequence of actions satisfying the following well-formedness conditions:*

1. *every action in $\tau$ has a unique identifier: if $\tau = \tau_1\,(a_1, _, _)\,\tau_2\,(a_2, _, _)\,\tau_3$ then $a_1 \neq a_2$;*
2. *commands in actions executed by a thread $t$ do not access local variables of other threads $t' \neq t$: if $\tau = _\,(_, t, c)\,_$ then $c \in \mathsf{LPcomm}_t$;*
3. *for every thread $t$, the projection $\tau|_t$ of $\tau$ onto the actions by $t$ cannot contain a request action immediately followed by a primitive action: if $\tau|_t = _\alpha_1\alpha_2_$ and $\alpha_1$ is a request then $\alpha_2$ is a response;*
4. *request and response actions are properly matched: for every thread $t$, $\mathsf{history}(\tau)|_t$ consists of alternating request and corresponding response actions, starting from a request action;*
5. *actions denoting the beginning and end of transactions are properly matched: for every thread $t$, in the projection of $\tau|_t$ to $\mathsf{begintx}$, $\mathsf{committed}$ and $\mathsf{aborted}$ actions, $\mathsf{begintx}$ alternates with $\mathsf{committed}$ or $\mathsf{aborted}$, starting from $\mathsf{begintx}$;*
6. *non-transactional accesses execute atomically: if $\tau = \tau_1\,\alpha\,\tau_2$, where $\alpha$ is a $\mathsf{read}$ or a $\mathsf{write}$ request action by thread $t$, and all the transactions of $t$ in $\tau_1$ completed, then $\tau_2$ begins with a response to $\alpha$.*
7. *non-transactional accesses never abort: if $\tau = _\,\alpha_1\,\alpha_2\,\tau_2$, where $\alpha_1$ is a non-transactional request action then $\alpha_2$ is not an $\mathsf{aborted}$ action.*

## A.2 Formal Definition of the Programming Language Semantics

The *semantics* of the programming language is the set of traces that computations of programs produce. We first describe its high-level structure, and then present its formalization. A *state* of a program $P = C_1 \parallel \ldots \parallel C_N$ records the values of all its variables: $s \in \mathsf{State} = (\biguplus_{t=1}^{N} \mathsf{LVar}_t) \to \mathbb{Z}$. The semantics of a program $P$ is given by the set of traces $[\![P, \mathcal{H}]\!](s) \subseteq \mathsf{Traces}$ it produces when executed with a TM $\mathcal{H}$ from an initial state $s$. To define this set, we first define the set of traces $[\![P]\!](s) \subseteq \mathsf{Traces}$ that a program can produce when executed from $s$ with the behavior of the TM unrestricted, i.e., considering all possible values the TM can return on reads and allowing transactions to commit or abort arbitrarily. This definition follows the intuitive semantics of our programming language. We then restrict $[\![P]\!](s)$ to the set of traces produced by $P$ when executed with $\mathcal{H}$ by selecting those traces that interact with the TM in a way consistent with $\mathcal{H}$: $[\![P, \mathcal{H}]\!](s) = \{\tau \mid \tau \in [\![P]\!](s) \wedge \mathsf{history}(\tau) \in \mathcal{H}\}$, where $\mathsf{history}(\cdot)$ projects to interface actions.

We now formally define the set $[\![P]\!](s)$. It is computed in two stages. First, we compute a set $A(P)$ of traces that resolves all issues regarding sequential control flow and interleaving. Intuitively, if one thinks of each thread $C_t$ in $P$ as a control-flow graph, then $A(P)$ contains all possible interleavings of paths in the graphs of $C_t$, $t \in \mathsf{ThreadID}$ starting from their initial nodes. The set $A(P)$ is a superset of all the traces that can actually be executed: e.g., if a thread executes the command "$x := 1$; `if` $(x = 1)$ $y := 1$ `else` $y := 2$" where $x, y$ are local variables, then $A(P)$ will contain a trace where $y := 2$ is executed instead of $y := 1$. To filter out such nonsensical traces, we *evaluate* every trace to determine whether it is ***valid***, i.e., whether its control flow is consistent with the effect of its actions on program variables. This is formalized by a function $\mathsf{eval} : \mathsf{State} \times \mathsf{Traces} \to \mathcal{P}(\mathsf{State}) \cup \{\lightning\}$ that, given an initial state and a trace, produces the set of states resulting from executing the actions in the trace, an empty set if the trace is invalid, or a special state $\lightning$ if the trace contains a `fault` action. Thus, $[\![P]\!](s) = \{\tau \in A(P) \mid \mathsf{eval}(s, \tau) \neq \emptyset\}$.

When defining the semantics, we encode the evaluation of conditions in `if` and `while` statements with `assume` commands. More specifically, we expect that the sets $\mathsf{LPcomm}_t$ contain special primitive commands $\mathtt{assume}(b)$, where $b$ is a Boolean expression over local variables of thread $t$, defining the condition. We state their semantics formally below; informally, $\mathtt{assume}(b)$ does nothing if $b$ holds in the current program state, and stops the computation otherwise. Thus, it allows the computation to proceed only if $b$ holds. The `assume` commands are only used in defining the semantics of the programming language; hence, we forbid threads from using them directly.

**The trace set** $A(P)$**.** The function $A'(\cdot)$ in Figure 4 maps commands and programs to sequences of actions they may produce. Technically, $A'(\cdot)$ might contain sequences that are not traces, e.g., because they do not have unique identifiers or continue beyond a `fault` command. This is resolved by intersecting the set $A'(P)$ with the set of all traces to define $A(P)$. $A'(C)t$ gives the set of action sequences produced by a command $C$ when it is executed by thread $t$. To define $A'(P)$, we first compute the set of all the interleavings of action sequences produced by the threads constituting $P$. Formally, $\tau \in \mathsf{interleave}(\tau_1, \ldots, \tau_m)$ if and only if every action in $\tau$ is performed by some thread $t \in \{1, \ldots, m\}$, and $\tau|_t = \tau_t$ for every thread $t \in \{1, \ldots, m\}$. We then let $A'(P)$ be the set of all prefixes of the resulting sequences which respect Definition 27, as denoted by the $\mathsf{prefix}$ operator. We take prefix closure here (while respecting the atomicity of non transactional access) to account for incomplete program computations as well as those in which the scheduler preempts a thread forever.

$A'(c)t$ returns a singleton set with the action corresponding to the primitive command $c$ (primitive commands execute atomically). $A'(C_1; C_2)t$ concatenates all possible action sequences corresponding to $C_1$ with those corresponding to $C_2$. The set of action sequences of a conditional considers cases where either branch is taken. We record the decision using an `assume` action; at the evaluation stage, this allows us to ensure that this decision is consistent with the program state. The set of action sequences for a loop is defined by considering all possible unfoldings of the loop body. Again, we record branching decisions using `assume` actions.

The set of action sequences of `read` and `write` accesses includes both sequences where the access executes successfully and where the current transaction is aborted. The former set is constructed by nondeterministically choosing an integers $v$ to describe the the return and parameter for the `read` and `write` accesses, respectively. To ensure that $e$ indeed evaluates to $v$, in the case of a `write`, Note that some of the choices here might not be feasible: the chosen $v$ might not be the value of the parameter expression $e$ when the method is invoked. Such infeasible choices are filtered out at the following stages of the semantics definition: the former in the definition of $[\![P]\!](s)$ by the use of evaluation and the semantics of `assume`, and the latter in the definition of $[\![P]\!](\mathcal{H}, s)$ by selecting the sequences from $[\![P]\!](s)$ that interact with the transactional memory correctly. The set of action sequences of $x := \mathtt{atomic}\,\{C\}$ contains those in which $C$ is aborted in the middle of its execution (at an object operation or right after it begins) and those in which $C$ executes until completion and then the transaction

$$
\begin{aligned}
A'(c)t &= \{(_, t, c)\}\\
A'(C_1; C_2)t &= \{\tau_1\,\tau_2 \mid \tau_1 \in A'(C_1)t \wedge \tau_2 \in A'(C_2)t\}\\
A'(\texttt{if}\,(b)\ \texttt{then}\ C_1\ \texttt{else}\ C_2)t &= \{(_, t, \texttt{assume}(b))\,\tau_1 \mid \tau_1 \in A'(C_1)t\} \cup \{(_, t, \texttt{assume}(\neg b))\,\tau_2 \mid \tau_2 \in A'(C_2)t\\
A'(\texttt{while}\,(b)\ \texttt{do}\ C)t &= \{\tau_1\,\tau_2 \ldots \tau_{2n}\,(_, t, \texttt{assume}(\neg b)) \mid n \in \mathbb{N} \wedge \forall j \in \{1, \ldots, n\}.\, \tau_{2j-1} = (_, t, \texttt{assume}(b))\\
&\qquad\qquad \wedge\, \tau_{2j} \in A'(C)t\} \cup \{(_, t, \texttt{assume}(\neg b))\}\\
A'(l := x.\texttt{read}())t &= \{(_, t, \mathsf{read}(x))\,(_, t, \mathsf{ret}(v))\,(_, t, l := v) \mid v \in \mathbb{Z}\} \cup \{(_, t, \mathsf{read}(x))\,(_, t, \mathsf{aborted})\}\\
A'(x.\texttt{write}(e))t &= \{(_, t, \texttt{assume}(e = v))\,(_, t, \mathsf{write}(x, v))\,(_, t, \mathsf{ret}(\bot))) \mid v \in \mathbb{Z}\}\\
&\quad \cup \{(_, t, \texttt{assume}(e = v))\,(_, t, \mathsf{write}(x, v))\,(_, t, \mathsf{aborted}) \mid v \in \mathbb{Z}\}\\
A'(x := \texttt{atomic}\,\{C\})t &= \{(_, t, \mathsf{begintx})\,(_, t, \mathsf{aborted})\,(_, t, x := \mathsf{aborted})\}\\
&\quad \cup \{(_, t, \mathsf{begintx})\,(_, t, \mathsf{ok})\,\tau\,(_, t, \mathsf{aborted})\,(_, t, x := \mathsf{aborted}) \mid\\
&\qquad\qquad \tau\,(_, t, \mathsf{aborted})\,\tau' \in A'(C)t \wedge (_, t, \mathsf{aborted}) \notin \tau\}\\
&\quad \cup \{(_, t, \mathsf{begintx})\,(_, t, \mathsf{ok})\,\tau\,(_, t, \mathsf{trycommit})\,(_, t, r)\,(_, t, x := r) \mid\\
&\qquad\qquad \tau \in A'(C)t \wedge (_, t, \mathsf{aborted}) \notin \tau \wedge (r = \mathsf{committed} \vee r = \mathsf{aborted})\}\\
A'(C_1 \parallel \ldots \parallel C_m) &= \mathsf{prefix}(\textstyle\bigcup\{\mathsf{interleave}(\tau_1, \ldots, \tau_m) \mid \forall t.\, 1 \le t \le m \implies \tau_t \in A'(C_t)t\})\\
A(P) &= A'(P) \cap \mathsf{Traces}
\end{aligned}
$$

**Figure 4** The definition of $A(P)$.

commits or aborts.

**Semantics of primitive commands.** To define evaluation, we assume a semantics of every command $c \in \mathsf{PCom}$, given by a function $[\![c]\!]$ that defines how the program state is transformed by executing $c$. As we noted before, different classes of primitive commands are supposed to access only certain subsets of variables. To ensure that this is indeed the case, we define $[\![c]\!]$ as a function of only those variables that $c$ is allowed to access. Namely, the semantics of $c \in \mathsf{LPcomm}_t$ is given by

$$[\![c]\!] : (\mathsf{LVar}_t \to \mathbb{Z}) \to \mathcal{P}(\mathsf{LVar}_t \to \mathbb{Z}).$$

Note that we allow $c$ to be non-deterministic.

For a valuation $q$ of variables that $c$ is allowed to access, $[\![c]\!](q)$ yields the set of their valuations that can be obtained by executing $c$ from a state with variable values $q$. For example, an assignment command $l := l'$ has the following semantics:

$$[\![l := l']\!](q) = \{q[l \mapsto q(l')]\}.$$

We define the semantics of `assume` commands following the informal explanation given at the beginning of this section: for example,

$$[\![\mathtt{assume}(l = v)]\!](q) = \begin{cases} \{q\}, & \text{if } q(l) = v; \\ \emptyset, & \text{otherwise.} \end{cases} \tag{1}$$

Thus, when the condition in `assume` does not hold of $q$, the command stops the computation by not producing any output.

We lift functions $[\![c]\!]$ to full states by keeping the variables that $c$ is not allowed to access unmodified and producing $\lightning$ if $c$ faults. For example, if $c \in \mathsf{LPcomm}_t$, then

$$[\![c]\!](s) = \{s|_{\mathsf{LVar} \setminus \mathsf{LVar}_t} \uplus q \mid q \in [\![c]\!](s|_{\mathsf{LVar}_t})\},$$

where $s|_V$ is the restriction of $s$ to variables in $V$. (For simplicity, we assume commands to not fault.)

**Trace evaluation.** Using the semantics of primitive commands, we first define the evaluation of a single action on a given state:

$$\begin{aligned} &\mathsf{eval} : \mathsf{State} \times \mathsf{Action} \to \mathcal{P}(\mathsf{State}) \\ &\mathsf{eval}(s, (_, t, c)) \;=\; [\![c]\!](s); \\ &\mathsf{eval}(s, \psi) \;=\; \{s\}. \end{aligned}$$

Note that this does not change the state $s$ as a result of TM interface, since their return values are assigned to local variables by separate actions introduced when generating $A(P)$.

We then lift $\mathsf{eval}$ to traces as follows:

$$\mathsf{eval} : \mathsf{State} \times \mathsf{Traces} \to \mathcal{P}(\mathsf{State})$$

$$\mathsf{eval}(s, \tau) = \begin{cases} \emptyset, & \text{if } \tau = \tau'\varphi \wedge \mathsf{eval}(s, \tau') = \emptyset; \\ \mathsf{evalna}(s, \tau|_{\neg\mathsf{abortact}}), & \text{otherwise,} \end{cases}$$

where $\tau|_{\neg\mathsf{abortact}}$ denotes the trace obtained from $\tau$ by removing all actions inside aborted transactions, and

$$\mathsf{evalna}(s, \tau) = \begin{cases} \{s\}, & \text{if } \tau = \varepsilon; \\ \{s'' \in \mathsf{eval}(s', \varphi) \mid s' \in \mathsf{evalna}(s, \tau')\}, & \text{if } \tau = \tau'\varphi. \end{cases}$$

The set of states resulting from evaluating trace $\tau$ from state $s$ is effectively computed by the helper function $\mathsf{evalna}(s, \tau)$, which ignores actions inside aborted transactions to model local variable roll-back. However, ignoring the contents of aborted transactions completely poses a risk that we might consider traces including sequences of actions inside aborted transactions that yield an empty set of states. To mitigate this, $\mathsf{eval}(s, \tau)$ recursively evaluates every prefix of $\tau$, thus ensuring that sequences of actions inside aborted transaction are valid.

Recall that we define $[\![P]\!](s)$ as the set of those traces from $A(P)$ that can be evaluated from $s$ without getting stuck, as formalized by eval. Note that this definition enables the semantics of `assume` defined by (1) to filter out traces that make branching decisions inconsistent with the program state. For example, consider again the program "$l := 1$; `if` $(l = 1)$ $l' := 1$ `else` $l' := 2$". The set $A(P)$ includes traces where both branches are explored. However, due to the semantics of the `assume` actions added to the traces according to Figure 4, only the trace executing $l' := 1$ will result in a nonempty set of final states after the evaluation and, therefore, only this trace will be included into $[\![P]\!](s)$.

## A.3 The Atomic TM

We define an idealized *atomic* TM $\mathcal{H}_{\mathsf{atomic}}$ where the execution of transactions does not interleave with that of other transactions or with non-transactional accesses. By instantiating the semantics of $[\![P, \mathcal{H}]\!](s)$ with this TM, we formalize the strongly atomic semantics [8] (transactional sequential consistency [11]).

The atomic TM $\mathcal{H}_{\mathsf{atomic}}$ contains only histories that are *non-interleaved*, i.e., where actions by one transaction do not overlap with the actions of another transaction or of non-transactional accesses. Note that by definition actions pertaining to different non-transactional accesses cannot interleave. Note also that transactions in a non-interleaved history do not have to be complete, because they may be produced by programs in our language, e.g., due to a non-terminating loop inside an atomic block. For example,

$$\begin{aligned} H_0 = {} & (_, t_1, \mathsf{begintx})\,(_, t_1, \mathsf{ok})\,(_, t_1, \mathsf{write}(x, 1))\,(_, t_1, \mathsf{ret}(\bot))\,(_, t_1, \mathsf{trycommit}) \\ & (_, t_2, \mathsf{begintx})\,(_, t_2, \mathsf{ok})\,(_, t_2, \mathsf{write}(x, 2))\,(_, t_3, \mathsf{read}(x))\,(_, t_3, \mathsf{ret}(1)) \end{aligned}$$

is non-interleaved. We have to allow such histories in $\mathcal{H}_{\mathsf{atomic}}$, because they may be produced by programs in our language, e.g., due to a non-terminating loop inside an atomic block.

We define $\mathcal{H}_{\mathsf{atomic}}$ in such a way that the changes made by a live or aborted transaction are invisible to other transactions. However, there is no such certainty in the treatment of a commit-pending transaction: the TM implementation might have already reached a point at which it is decided that the transaction will commit. Then the transaction is effectively committed, and its operations may affect other transactions [21]. To account for this, when defining $\mathcal{H}_{\mathsf{atomic}}$ we consider every possible completion of each commit-pending transaction in a history to either committed or an aborted one. Formally, we say that a history $H^c$ is a *completion* of a non-interleaved history $H$ if:

1. $H^c$ is non-interleaved;
2. $H^c$ is has no commit-pending transactions;
3. $H$ is a subsequence of $H^c$; and
4. any action in $H^c$ which is not in $H$ is either a committed or an aborted action.

For example, we can obtain a completion of history $H_0$ above by inserting $(_, t_1, \mathsf{committed})$ after $(_, t_1, \mathsf{trycommit})$.

We define $\mathcal{H}_{\mathsf{atomic}}$ as the set of all non-interleaved histories $H$, in which every transaction can be completed that have a completion $H^c$ where every response action of a $\mathsf{read}(x)$ returns the value $v$ in the last preceding $\mathsf{write}(x, v)$ action that is not located in an aborted or live transaction different from the one of the $\mathsf{read}$; if there is no such $\mathsf{write}$, the read should return the initial value $v_{\mathsf{init}}$. For example, $H_0 \in \mathcal{H}_{\mathsf{atomic}}$. Hence, $\mathcal{H}_{\mathsf{atomic}}$ defines the intuitive atomic semantics of transactions.

## B Additional Details for §3: Examples of Privatization

$$\{\, \texttt{x_is_ready} = \texttt{false} \wedge \texttt{x} = 0 \,\}$$

```
l1 = atomic {                  ||  do {
  x = 42;                      ||    l2 = x_is_ready; // n'
} // T                         ||  } while (¬l2);
x_is_ready = true; // n        ||  l3 = x; // n''
```

$$\{\, \texttt{l}_1 = \texttt{committed} \implies \texttt{l}_3 = 42 \,\}$$

**Figure 5** Privatization by agreement outside transactions.

$$\{\, \texttt{priv} = \texttt{false} \wedge \texttt{x} = 0 \,\}$$

```
atomic {          ||  l1 = atomic {                ||  atomic {
  priv = true;    ||    l2 = priv;                 ||    if (¬priv)
} // T1           ||  } // T2                      ||      x = 42;
                  ||  if (l1==committed ∧ l2)      ||  } // T3
                  ||    x = 1; // n                ||
```

$$\{\, \texttt{l}_1 = \texttt{committed} \wedge \texttt{l}_2 \implies \texttt{x} = 1 \,\}$$

**Figure 6** Proxy privatization.

Figure 5 gives an example of a program that privatizes an object by agreeing on its status outside transactions ("partitioning by consensus" in [34]). The left-hand-side thread writes to `x` inside a transaction and then sets the flag `x_is_ready` outside. The right-hand-side thread keeps reading the flag non-transactionally until it is set, and then reads `x` non-transactionally. Since the program is TDRF, we expect the postcondition shown to hold: . This program is TDRF because, in any of its traces under $\mathcal{H}_{\mathsf{atomic}}$, the conflicting write in $T$ and the non-transactional read $n''$ are ordered in happens-before due to the client order between the write in $n$ and the read in $n'$ that causes the `do` loop to terminate. Since the program is TDRF, we expect the postcondition shown to hold.

Figure 6 gives an example of *proxy privatization* [36]. Analogously to the example from Figure 1, an object `x` is guarded by a flag `priv`, showing whether the object should be accessed transactionally (`false`) or non-transactionally (`true`). The left-hand-side thread first tries to set the flag inside transaction $T_1$. The middle and the right-hand-side threads both check the flag `priv` prior to accessing `x`, non-transactionally and transactionally, respectively. This program is TDRF because, in any of its traces under $\mathcal{H}_{\mathsf{atomic}}$, the non-transactional write $n$ and the conflicting write in $T_3$ are ordered in happens-before. Indeed, in every corresponding trace under $\mathcal{H}_{\mathsf{atomic}}$, $T_3$ executes before $T_1$ to justify reading `priv` being `false`, and $T_1$ executes before $T_2$ to enable the condition for the subsequent non-transactional access $n$. Since the program is TDRF, we expect the postcondition shown to hold.

## C Additional Details for §4: Proof of Lemma 7

We use some some of auxiliary lemmas, adapted from [6]. The following lemma shows that a trace $\tau_H$ with a history $H$ can be transformed into an equivalent trace $\tau_S$ with a history $S$ that is in the opacity relation with $H$.

▶ **Lemma 28.**

$$\forall H, S \in \mathsf{History}.\, H \sqsubseteq S \implies (\forall \tau_H.\, \mathsf{history}(\tau_H) = H \implies \exists \tau_S.\, \mathsf{history}(\tau_S) = S \wedge \tau_H \sim \tau_S).$$

We also rely on the following proposition that allows us to conclude that the trace $\tau_S$ resulting from the rearrangement in Lemma 28 can be produced by a program $P$ if so can the original trace $\tau_H$.

▶ **Proposition 29.** *If $\tau_H \in [\![P]\!](s)$ and $\tau_H \sim \tau_S$, then $\tau_S \in [\![P]\!](s)$.*

**Proof of Lemma 7.** Let us consider any program $P$ and assume that $\mathsf{TDRF}(P)$ holds. By Definition 3, we have:

$$\forall \tau \in [\![P]\!](\mathcal{H}_{\mathsf{atomic}}).\, \mathsf{TDRF}(\mathsf{history}(\tau)).$$

We consider any trace $\tau_H \in [\![P]\!](\mathcal{H})$ and its history $H = \mathsf{history}(\tau_H)$. We also consider any history $S \in \mathcal{H}_{\mathsf{atomic}}$ such that $H \sqsubseteq S$ holds. By Lemma 28, there exists $\tau_S$ such that $\mathsf{history}(\tau_S) = S$ and $\tau_H \sim \tau_S$. By Proposition 29, $\tau_S \in [\![P]\!](s)$ holds. Moreover, since $\mathsf{history}(\tau_S) = S \in \mathcal{H}_{\mathsf{atomic}}$, it is in fact the case that $\tau_S \in [\![P]\!](\mathcal{H}_{\mathsf{atomic}}, s)$.

Since the program $P$ is TDRF, histories of every trace from $[\![P]\!](\mathcal{H}_{\mathsf{atomic}})$ are TDRF too. In particular, $\mathsf{TDRF}(S)$ holds, which allows us to state the following:

$$\forall S.\, S \in \mathcal{H}_{\mathsf{atomic}} \wedge H \sqsubseteq S \implies \mathsf{TDRF}(S).$$

Hence, $\mathsf{CDRF}(H)$ holds by Definition 5. Overall, for every trace $\tau_H \in [\![P]\!](\mathcal{H}, s)$ we proved that $\mathsf{CDRF}(\mathsf{history}(\tau_H))$ holds. The latter allows us to conclude that $\mathsf{CDRF}(P, \mathcal{H})$ holds. ◀

# D Additional Details for §6: Proving Privatization-Safe Opacity

## D.1 Consistent Histories

▶ **Definition 30.** *A pair of matching request and response actions* $(\alpha, \alpha')$ *is* local *to a given transaction* $T$ *from a history* $H$*, if:*

- $\alpha = (_, _, \mathsf{read}(x)) \wedge \exists \beta \in T.\, \beta <_{\mathsf{po}(H)} \alpha \wedge \beta = (_, _, \mathsf{write}(x, _))$*; or*
- $\alpha = (_, _, \mathsf{write}(x, _)) \wedge \exists \beta \in T.\, \alpha <_{\mathsf{po}(H)} \beta \wedge \beta = (_, _, \mathsf{write}(x, _))$*.*

*We let* $\mathsf{local}(H)$ *denote the set of all local actions in* $H$*.*

▶ **Definition 31.** *We say that a pair* $(\alpha, \alpha')$ *is a* well-formed read *from* $x \in \mathsf{Reg}$ *in a history* $H$*, written* $\mathsf{wfWR}_{(H,x)}(\alpha, \alpha')$*, if* $\alpha = (_, _, \mathsf{write}(x, v))$*,* $\alpha' = (_, _, \mathsf{ret}(v))$ *and the matching request action for* $\alpha'$ *is* $(_, _, \mathsf{read}(x))$*.*

▶ **Definition 32.** *In a history* $H$*, a read request* $\alpha = (_, _, \mathsf{read}(x))$ *and its matching response* $\alpha' = (_, _, \mathsf{ret}(v))$ *are* consistent*, if:*

- *when* $(\alpha, \alpha') \in \mathsf{local}(H)$ *and performed by a transaction* $T$*,* $v$ *is the value written by the most recent write* $(_, _, \mathsf{write}(x, v))$ *preceding the read in* $T$*;*
- *when* $(\alpha, \alpha') \notin \mathsf{local}(H)$*, either there exists a non-local* $\beta$ *not located in an aborted or live transaction such that* $\mathsf{wfWR}_{(H,x)}(\beta, \alpha')$ *holds, or there is no such* $\beta$ *and* $v = v_{\mathsf{init}}$*.*

*Also, a history* $H$ *is* consistent*, written* $\mathsf{cons}(H)$*, if all of its matching read requests and responses are.*

## D.2 Proof of Proposition 18

We only prove the part 1, as part 2 can be proven analogously. Let $\mathsf{HB}_k(H) \triangleq (\mathsf{PO}(H) \cup \mathsf{CL}(H) \cup \mathsf{EF}(H))^k$ and let $\Phi(k)$ denote the following statement: if $T <_{\mathsf{HB}_k(H)} n$, then there are $T'$ and $n'$ such that $T \leq_{\mathsf{EF}(H)} T' <_{\mathsf{PO}(H)} n' \leq_{\mathsf{CL}(H)} n$. We prove $(\forall k \geq 1.\, \Phi(k))$ by induction on $k$.

The base of the induction, $\Phi(1)$, is when $T <_{\mathsf{HB}_1(H)} n$ holds. Out of relations in $\mathsf{HB}_1(H)$, only $\mathsf{PO}(H)$ can relate a transaction and a non-transactional access, meaning that $T <_{\mathsf{PO}(H)} n$ is the case. Hence, there exist $T' = T$ and $n' = n$ such that $T \leq_{\mathsf{EF}(H)} T' <_{\mathsf{PO}(H)} n' \leq_{\mathsf{CL}(H)} n$.

For the induction step, we consider $k > 1$ and assume that $\Phi(k')$ holds of each $k'$ such that $1 \leq k' \leq k - 1$. Let us assume that $T <_{\mathsf{HB}_k(H)} n$ holds too. By definition of $\mathsf{HB}_k(H)$, one of the following is the case:

1. there exists $T''$ such that $T <_{\mathsf{HB}_1(H)} T'' <_{\mathsf{HB}_{k-1}(H)} n$, or
2. there exists $n''$ such that $T <_{\mathsf{HB}_1(H)} n'' <_{\mathsf{HB}_{k-1}(H)} n$.

In the first case, applying the induction hypothesis $\Phi(k-1)$ to $T'' <_{\mathsf{HB}_{k-1}(H)} n$ immediately concludes the proof. In the second case, by definition of $\mathsf{CL}$, $n'' <_{\mathsf{CL}(H)} n$ holds. Therefore, $T <_{\mathsf{HB}_2(H)} n$ holds, so we can apply the induction hypothesis $\Phi(2)$ to conclude the proof. ◀

## D.3 Proof of Theorem 20

▶ **Proposition 33.** *If a strict partial order* $R \subseteq Y \times Y$ *is acyclic and sets* $X, Z \subseteq Y$ *are such that:*

- $y \notin X$ *and* $y \notin Z$*,*
- $\forall x \in X.\, \neg(y \xrightarrow{R}{}^{*} x)$*,*
- $\forall z \in Z.\, \neg(z \xrightarrow{R}{}^{*} y)$*, and*
- $\forall x \in X, z \in Z.\, \neg(z \xrightarrow{R}{}^{*} x)$*,*

*then* $R \cup \{(x, y) \mid x \in X\} \cup \{(y, z) \mid z \in Z\}$ *is acyclic.*

**Proof.** We prove the proposition in two steps. First, we prove acyclicity of $R_1 \triangleq R \cup \{(x, y) \mid x \in X\}$, and then of $R_2 \triangleq R_1 \cup \{(y, z) \mid z \in Z\}$.

We prove acyclicity of $R_1$ by contradiction. Since $R$ is acyclic, a cycle in $R_1$ must involve at least one new edge. Let $x \in X$ be such that $x \xrightarrow{R_1} y \xrightarrow{R_1}{}^* x$. However, $y \xrightarrow{R}{}^* x$ does not hold by the premise of the proposition. Therefore, $y \xrightarrow{R_1}{}^* x$ must also involve new edges. Let $x'$ be the element involved in the last of such edges:

$$x \xrightarrow{R_1} y \xrightarrow{R_1}{}^* x' \xrightarrow{R_1} y \xrightarrow{R}{}^* x.$$

We arrived to a contradiction to $\neg(y \xrightarrow{R}{}^* x)$.

We now prove acyclicity of $R_2$. Let us show that the following holds:

$$\forall z \in Z. \neg(z \xrightarrow{R_1}{}^* y). \tag{2}$$

We do that by contradiction. Let us assume there is $z \in Z$ such that $z \xrightarrow{R_1}{}^* y$. By the premise of the proposition, $z \xrightarrow{R}{}^* y$ does not hold, so at least one new edge of $R_1$ must participate in $z \xrightarrow{R_1}{}^* y$. However, since all such edges end in $y$ and $R_1$ is acyclic, it can only be that for some $x \in X$, $z \xrightarrow{R}{}^* x \xrightarrow{R_1} y$ holds. The latter contradicts the premise of the proposition. Therefore, (2) holds. This observation enables demonstrating acyclicity of $R_2$ analogously to the first step of the proof. ◀

In the following lemma, we denote the set of non-transactional accesses in $\mathcal{V}$ by $\mathsf{nontxn}(\mathcal{V})$ and range over them by $n$. We also let $\mathsf{txns}(\mathcal{V})$ be the set of transactions in $\mathcal{V}$.

▶ **Lemma 34.** *Consider an acyclic opacity graph $G = (\mathcal{V}, \mathsf{vis}, \mathsf{WR}, \mathsf{WW}, \mathsf{RW}, \mathsf{PO}, \mathsf{CL})$ of a consistent CDRF history $H$. For any conflicting non-transactional access $n$ and a transaction $T$, either $n \xrightarrow{\mathsf{DEP}}{}^* T$ or $T \xrightarrow{\mathsf{DEP}}{}^* n$ holds.*

**Proof.** Let us consider any conflicting non-transactional access $n$ and a transaction $T$. We prove the lemma by contradiction. We assume that the following holds in the graph $G$:

$$\neg(n \xrightarrow{\mathsf{DEP}}{}^* T \vee T \xrightarrow{\mathsf{DEP}}{}^* n), \tag{3}$$

and show that this assumption contradicts CDRF of $H$.

We represent the set of linearizations of $G$, $\mathsf{lins}(G)$, as a disjoint union of two sets: $\mathcal{L}_1$, the set of all linearizations of $G$, in which $T$ occurs before $n$, and $\mathcal{L}_2$, all other linearizations of $G$. We then let $\mathcal{A}_1$ and $\mathcal{A}_2$ denote the following sets of transactions:

$$\mathcal{A}_1 \triangleq \{T_1 \mid \exists L_1, n_1. L_1 \in \mathcal{L}_1 \wedge n_1 \in \mathsf{nontxn}(\mathcal{V}) \wedge T \leq_{\mathsf{EF}(L_1)} T_1 <_{\mathsf{PO}(L_1)} n_1 \leq_{\mathsf{CL}(L_1)} n\};$$
$$\mathcal{A}_2 \triangleq \{T_2 \mid \exists L_2, n_2. L_2 \in \mathcal{L}_2 \wedge n_2 \in \mathsf{nontxn}(\mathcal{V}) \wedge n \leq_{\mathsf{CL}(L_2)} n_2 <_{\mathsf{PO}(L_2)} T_2 \leq_{\mathsf{EF}(L_2)} T\}.$$

With the help of Proposition 33, we demonstrate that history $H$ has a linearization in which every $T_1 \in \mathcal{A}_1$ occurs before $T$ and every $T_2 \in \mathcal{A}_2$ occurs after $T$. Consider any $T_1 \in \mathcal{A}_1$ and $T_2 \in \mathcal{A}_2$. By the definition of $T_1 \in \mathcal{A}_1$, there exist $L_1 \in \mathcal{L}_1$ and $n_1 \in \mathsf{nontxn}(\mathcal{V})$ such that $T_1 <_{\mathsf{PO}(L_1)} n_1 \leq_{\mathsf{CL}(L_1)} n$. By Definition 12 of opacity graphs, $T_1 \xrightarrow{\mathsf{DEP}}{}^* n$ holds of $G$. Analogously, $n \xrightarrow{\mathsf{DEP}}{}^* T_2$ holds. Since $G$ is acyclic and $T_1 \xrightarrow{\mathsf{DEP}}{}^* T_2$ holds, we can conclude the following:

$$\forall T_1 \in \mathcal{A}_1, T_2 \in \mathcal{A}_2. \neg(T_2 \xrightarrow{\mathsf{DEP}}{}^* T_1). \tag{4}$$

By assumption (3), neither $T \xrightarrow{\mathsf{DEP}}{}^{*} n$ nor $n \xrightarrow{\mathsf{DEP}}{}^{*} T$ holds. Therefore, $T \in \mathcal{A}_1$, $T \in \mathcal{A}_2$, $T \xrightarrow{\mathsf{DEP}}{}^{*} T_1$ and $T_2 \xrightarrow{\mathsf{DEP}}{}^{*} T$ do not hold either (otherwise, $T \xrightarrow{\mathsf{DEP}}{}^{*} n$ and $n \xrightarrow{\mathsf{DEP}}{}^{*} T$ would be implied by transitivity). Overall, we have shown:

$$\begin{aligned} &\forall T_1 \in \mathcal{A}_1.\, \neg(T \xrightarrow{\mathsf{DEP}}{}^{*} T_1);\\ &\forall T_2 \in \mathcal{A}_2.\, \neg(T_2 \xrightarrow{\mathsf{DEP}}{}^{*} T). \end{aligned} \tag{5}$$

By Proposition 33, (4) and (5) imply that the graph $G$ has a linearization in which every $T_1 \in \mathcal{A}_1$ occurs before $T$ and every $T_2 \in \mathcal{A}_2$ occurs after $T$.

In the rest of the proof, we consider a linearization $L \in \mathsf{lins}(G)$, in which every $T_1 \in \mathcal{A}_1$ occurs before $T$ and every $T_2 \in \mathcal{A}_2$ occurs after $T$. History $H$ is consistent and its opacity graph $G$ is acyclic. By Lemma 13, $L \in \mathsf{lins}(G) \subseteq \mathcal{H}_{\mathsf{atomic}}$ holds. Moreover, $H$ is CDRF, meaning that the conflicting pair of $T$ and $n$ must be ordered by $\mathsf{HB}(L)$, that is, either $T <_{\mathsf{HB}(L)} n$ or $n <_{\mathsf{HB}(L)} T$ holds. By Proposition 18, $T <_{\mathsf{HB}(L)} n$ holds only if there exist $T_1$ and $n_1$ such that:

$$T \leq_{\mathsf{EF}(L)} T_1 <_{\mathsf{PO}(L)} n_1 \leq_{\mathsf{CL}(L)} n. \tag{6}$$

Let us assume that such $T_1$ and $n_1$ exist. Then $T \leq_{\mathsf{EF}(L)} T_1$ holds. However, it is the case that $T_1 \in \mathcal{A}_1$, and $T_1 \leq_{\mathsf{EF}(L)} T$ must hold by construction of $L$. We arrived to a contradiction, meaning that there are no $T_1$ and $n_1$ satisfying (6). We conclude that $n <_{\mathsf{HB}(L)} T$ does not hold, and, analogously, we can show that $T <_{\mathsf{HB}(L)} n$ does not hold either. Overall, we have found a linearization $L$ of a CDRF history $H$ such that TDRF does not hold of $L$. We have arrived to a contradiction with CDRF of $H$. ◀

By combining Lemmas 16 and 34, we get

▶ **Corollary 35.** *Consider an acyclic opacity graph $G = (\mathcal{V}, \mathsf{vis}, \mathsf{WR}, \mathsf{WW}, \mathsf{RW}, \mathsf{PO}, \mathsf{CL})$ of a consistent CDRF history $H$. For any conflicting non-transactional access $n$ and a transaction $T$, either $n \xrightarrow{\mathsf{txDEP} \cup \mathsf{RT} \cup \mathsf{PO} \cup \mathsf{CL}}{}^{+} T$ or $T \xrightarrow{\mathsf{txDEP} \cup \mathsf{RT} \cup \mathsf{PO} \cup \mathsf{CL}}{}^{+} n$ holds.*

**Proof of Theorem 20.** The "if" case is covered by Corollary 35, so here we prove the "only if" case. Let us consider any $L \in \mathcal{H}_{\mathsf{atomic}}$ such that $H \sqsubseteq L$. According to the definition of $\mathcal{H}_{\mathsf{atomic}}$ (§A.3), $L$ has a non-interleaved completion $L'$. We construct an opacity graph $G' = (\mathcal{V}, \mathsf{vis}, \mathsf{WR}, \mathsf{WW}, \mathsf{RW}, \mathsf{PO}, \mathsf{CL})$ based on $L$ and $L'$ as follows:

- $\mathcal{V}$ is the set of all transactions and non-transactional accesses in $L$.
- $\mathsf{vis}$ is the set of committed transactions and all non-transactional accesses in $L'$.
- For each $x \in \mathsf{Reg}$, we let $\nu \xrightarrow{\mathsf{WR}_x} \nu'$ if $\nu'$ reads from $x$ and $\nu$ is the $<_L$-last out of visible vertexes writing to $x$ and preceding $\nu'$.
- For each $x \in \mathsf{Reg}$, we let $\mathsf{WW}_x$ be the order in which visible vertexes writing to $x$ occur in $L$.
- We derive $\mathsf{PO}$ and $\mathsf{CL}$ from $L$, and $\mathsf{RW}$ from $\mathsf{WR}$ and $\mathsf{WW}$.

It is easy to show that $G \in \mathsf{Graph}(H)$. Therefore, by the premise of the theorem, between every conflicting pair of a transaction $T$ and a non-transactional access $n$ there is a path in $(\mathsf{txDEP} \cup \mathsf{RT} \cup \mathsf{PO} \cup \mathsf{CL})^{+}$. Note that whenever $T_1 \xrightarrow{\mathsf{txDEP} \cup \mathsf{RT}} T_2$, $T_1 <_{\mathsf{EF}(L)} T_2$ holds. Consequently, $T$ and $n$ are ordered by $<_{\mathsf{HB}(L)}$. Overall, we have shown:

$$\forall L.\, (H \sqsubseteq L \land L \in \mathcal{H}_{\mathsf{atomic}}) \implies \mathsf{TDRF}(L),$$

which allows us to conclude $\mathsf{CDRF}(H)$. ◀

# E Additional Details for §7: Case Studies

## E.1 FencedTL2

In §7, as a part of the proof of Theorem 21 we showed that acyclicity of opacity graphs is invariant under the graph update $\textsc{txread}(T, x)$. We now give details for the other graph updates. Note that for the pseudo-code of TL2, we refer to [26, §C.1].

We make the following observation about the graph updates in FencedTL2: both transactional and non-transactional reads return the value of the most recent write to the object. Based on that, we state the following proposition.

▶ **Proposition 36.** *The graph updates* $\textsc{txread}(\nu, x)$ *and* $\textsc{ntxread}(\nu, x)$ *do not add edges of the form* $\nu \xrightarrow{\mathsf{RW}_x} _$.

The proposition holds trivially of non-transactional reads. For $\textsc{txread}(T, x)$, we note that at the moment of the graph update adding a read dependency $\nu \xrightarrow{\mathsf{WR}_x} T$, $\nu$ can be shown to be the most recent write to $x$ (this proof is a part of proving the usual opacity of TL2). Adding a read dependency on the most recent write does not induce anti-dependencies by Definition 12 of RW.

▶ **Lemma 37.** *Let* $(H', G')$ *be the result of performing an update* $\textsc{txinit}(T)$, $\textsc{ntxread}(n, x)$ *or* $\textsc{ntxwrite}(n, x)$ *on* $(H, G)$. *Assume that* $(H, G), (H', G') \in \mathsf{INV}$ *and* $G$ *is acyclic. Then* $G'$ *is acyclic too.*

**Proof.** The graph event $\textsc{txinit}(T)$ does not invalidate acyclicity of the graph, because it only adds a new vertex into the graph and orders it after some of the existing vertexes. This vertex does not have outgoing edges, so it cannot take part in a cycle.

The graph events $\textsc{ntxread}(n, x)$ and $\textsc{ntxwrite}(n, x)$ do not invalidate acyclicity of the graph either. They only add edges ending in the vertex $n$, i.e., they do not add edges of the form $n \xrightarrow{\mathsf{DEP}} _$. Also, upon the execution of these graph events, $n$ is a new vertex in the graph, and it does not have outgoing edges. Therefore, $n$ cannot take part in a cycle. ◀

▶ **Lemma 38.** *Let* $(H', G')$ *be the result of performing an update* $\textsc{txread}(T, x)$ *on* $(H, G)$. *Assume that* $(H, G), (H', G') \in \mathsf{INV}_1 \wedge \mathsf{INV}_2$ *and* $G$ *is acyclic. Then* $G'$ *is acyclic too.*

**Proof.** By contrapositive: we assume that $G'$ contains a simple cycle and show that $G'$ violates either $\mathsf{INV}_1$ or $\mathsf{INV}_2$. The graph update adds an edge of the form $_ \xrightarrow{\mathsf{WR}_x} T$. Consider a simple cycle involving the new edge $\nu \xrightarrow{\mathsf{WR}_x} T$ for some vertex $\nu$. By our assumption, there must be a reverse path $T \xrightarrow{\mathsf{DEP}}{}^* \nu$ in $G$. Let us first consider the case when $\nu$ is a transaction $T'$. Since $G$ is acyclic and $H$ is consistent and CDRF, by Lemma 15 the path $T \xrightarrow{\mathsf{DEP}}{}^* T'$ can be reduced to $T \xrightarrow{\mathsf{RT} \cup \mathsf{txDEP}}{}^* T'$. Since $G'$ only extends $G$, the same path is present in $G'$ too. Then $T' \xrightarrow{\mathsf{txDEP}} T \xrightarrow{\mathsf{RT} \cup \mathsf{txDEP}}{}^* T'$ is a cycle over transactions in $G'$, which contradicts $(H', G') \in \mathsf{INV}_1$.

We now consider the case when $\nu$ is a non-transactional access $n$. Since $G$ is acyclic and $H$ is consistent and CDRF, by Lemma 16 there exist $T'$ and $n'$ such that $T \xrightarrow{\mathsf{txDEP}}{}^* T' \xrightarrow{\mathsf{PO}} n' \xrightarrow{\mathsf{CL}}{}^* n$ holds of $G$. Note that $T$ is an uncompleted transaction, since it currently performs a graph update. Therefore, $T \xrightarrow{\mathsf{txDEP}}{}^* T' \xrightarrow{\mathsf{PO}} n' \xrightarrow{\mathsf{CL}}{}^* n$ is a contradiction to $(H, G) \in \mathsf{INV}_2$.

◀

## E.2 Two-Phase Locking TM

In this section, we consider a two-phase locking TM system [21], which we call 2PL. This TM assumes a try-lock for every register in memory, and synchronizes transactional accesses to the registers according to the two-phase locking protocol. 2PL also maintains transaction-local read- and write-sets similarly to other TMs. To perform a transactional read from a register $x$ in a transaction $T$, 2PL first checks if $x$ is in either read- or write-set of $T$, in which case $T$ returns the previously read or written value accordingly, and otherwise $T$ attempts to acquire a lock on $x$. After having successfully acquired the latter, $T$ reads the value of $x$ from memory and adds a corresponding record into the read-set. To perform a transactional write to a register $x$ in a transaction $T$, 2PL first checks $x$ is not in the write-set of $T$ yet, in which case $T$ tries to acquire a lock on $x$, modifies the register in memory and stores a record in the write-set (together with the original value of $x$). If $T$ already has $x$ in its write-set, the transaction simply writes to memory. To commit a transaction, 2PL simply releases all the locks it has acquired, and to abort one, 2PL first rolls back the updates to registers by writing back to memory their original values as stored in the write-set, and then releases the locks.

To prove privatization-safe opacity of 2PL, for every one of its executions we inductively construct an opacity graph matching its history with the help of the following *graph updates*, which specify how and when in the execution to extend the graph:

- At the start of a transaction $T$, a graph update $\textsc{txinit}(T)$ adds a new vertex $T$ and extends the real-time order with edges $T' \xrightarrow{\mathsf{RT}} T$ for every completed transaction $T'$.
- At the end of a read operation of a transaction $T$ reading from an object $x$, a graph update $\textsc{txread}(T, x)$ adds a read dependency $\nu \xrightarrow{\mathsf{WR}_x} T$, where $\nu$ is the vertex that wrote the value returned by the read.
- Upon starting the commit of a transaction $T$, a graph update $\textsc{txwrite}(T, x)$ adds a write dependency $\nu \xrightarrow{\mathsf{WW}_x} T$ for every object $x$ in the write-set of $T$, where $\nu$ is the vertex that wrote the previous value of $x$.
- Upon each non-transactional write $n$ to an object $x$, a graph update $\textsc{ntxwrite}(n, x)$ adds a new vertex $n$ and a write-dependency $\nu \xrightarrow{\mathsf{WW}_x} n$, where $\nu$ is the vertex that wrote the previous value of $x$.
- Upon each non-transactional read $n$ from an object $x$, a graph update $\textsc{ntxread}(n, x)$ adds a new vertex $n$ and a read dependency $\nu \xrightarrow{\mathsf{WR}_x} n$, where $\nu$ is the vertex that wrote the value returned by $n$.

The updates also add anti-dependencies of the form $_ \xrightarrow{\mathsf{RW}} T$ induced by new read- and write-dependencies.

At each step of the graph construction we prove that the graph remains acyclic. Then Theorem 14 guarantees that the history of the execution is opaque. We use Lemmas 15 and 16 to reduce the task of proving the graph acyclicity to proving the absence of cycles involving transactions only.

We establish the following invariants over pairs $(H, G)$ of a history $H$ and a graph $G$:

- $\mathsf{INV}_1$. $H$ is a consistent history and the relation $\mathsf{txDEP} \cup \mathsf{RT}$ is acyclic.

To enable the reduction from privatization-safe to plain opacity, we prove the following invariant, which states the guarantee provided by the two-phase locking protocol in 2PL:

- $\mathsf{INV}_2$. For every uncompleted transaction $T$ such that its last action in $H$ is a response, and a transaction $T'$, $T \xrightarrow{\mathsf{txDEP}} T'$ does not hold.

An informal justification of the invariant is as follows. By construction of the graph it is possible to establish that a dependency $T \xrightarrow{\mathsf{txDEP}} T'$ can only be added by $\textsc{txread}(T', _)$ and $\textsc{txwrite}(T', _)$. However, to perform those updates, $T'$ first needs to acquire corresponding locks, which $T$ holds until it starts committing or aborting. In both cases, the last action by $T$ in the history $H$ would need to be a request (either for an operation or a commit).

By Theorem 14, privatization-safe opacity of 2PL follows from the following theorem.

▶ **Theorem 39.** $\forall H \in \text{2PL}.\, \mathsf{CDRF}(H) \implies \exists G.\,(H, G) \in \mathsf{INV}_1 \wedge \mathsf{INV}_2 \wedge \mathsf{acyclic}(G)$.

We prove Theorem 39 by induction on the length of the TM execution inducing $H$, constructing $G$ as described above and showing that it remains acyclic after each update with the aid of the two invariants. In this section we only explain how we prove acyclicity in the case of graph update $\textsc{txread}$ $\textsc{txwrite}$, which illustrates the use of Lemmas 15 and 16.

▶ **Lemma 40.** *When $(s, H, G) \in \mathsf{INV}$ and $(s', H', G')$ is the result of executing $\textsc{txinit}(T)$, $\textsc{ntxread}(n, x)$ and $\textsc{ntxwrite}(n, x)$, the graph $G'$ is acyclic.*

*Let $(H', G')$ be the result of performing an update $\textsc{txinit}(T)$, $\textsc{ntxread}(n, x)$ or $\textsc{ntxwrite}(n, x)$ on $(H, G)$. Assume that $(H, G), (H', G') \in \mathsf{INV}$ and $G$ is acyclic. Then $G'$ is acyclic too.*

**Proof.** The graph event $\textsc{txinit}(T)$ does not invalidate acyclicity of the graph, because it only adds a new vertex into the graph and orders it after some of the existing vertexes. This vertex does not have outgoing edges, so it cannot take part in a cycle.

The graph events $\textsc{ntxread}(n, x)$ and $\textsc{ntxwrite}(n, x)$ do not invalidate acyclicity of the graph either. They only add edges ending in the vertex $n$, i.e., they do not add edges of the form $n \xrightarrow{\mathsf{DEP}} _$. Also, upon the execution of these graph events, $n$ is a new vertex in the graph, and it does not have outgoing edges. Therefore, $n$ cannot take part in a cycle. ◀

The following proposition states a simple property of 2PL, which allows us to conclude that the graph updates of a vertex $\nu$ only add edges ordering $\nu$ after other vertexes.

▶ **Proposition 41.** *The graph updates $\textsc{txread}(\nu, x)$ and $\textsc{ntxread}(\nu, x)$ do not add edges of the form $\nu \xrightarrow{\mathsf{RW}_x} _$.*

The proposition can be proven inductively based on the observation that reads always return values written by most recent writes.

▶ **Lemma 42.** *Let $(H', G')$ be the result of performing an update $\textsc{txwrite}(T, x)$ on $(H, G)$. Assume that $(H, G), (H', G') \in \mathsf{INV}$ and $G$ is acyclic. Then $G'$ is acyclic too.*

**Proof.** By contrapositive: we assume that $G'$ contains a simple cycle and show that $G'$ violates $\mathsf{INV}$. The graph update only adds edges of the form $_ \xrightarrow{\mathsf{DEP}_x} T$ (in particular, Proposition 41 rules out a possibility of new anti-dependencies). Since all new edges end in the same vertex $T$, they cannot occur in the same simple cycle. Hence, we can consider them separately.

Consider a simple cycle involving a new edge $\nu \xrightarrow{\mathsf{DEP}} T$ for some vertex $\nu$. By our assumption, there must be a reverse path $T \xrightarrow{\mathsf{DEP}}{}^{*} \nu$ in $G$. Let $\nu'$ be the first action on the path after $T$. Note that $T$ is an uncompleted transaction in $(H, G)$, since it currently performs a graph update, and the last action by $T$ in $H$ is a response.

Thus, if $\nu'$ is some transaction $T'$, we obtain a contradiction to $\mathsf{INV}_2$. Let us now consider the case when $\nu'$ is a non-transactional access $n$. Since $G$ is acyclic and $H$ is consistent and CDRF, and the edge $T \xrightarrow{\mathsf{DEP}}{}^{*} \nu$ is in $G$, by Lemma 16 there exist $T'$ and $n'$ such that $T \xrightarrow{\mathsf{txDEP}}{}^{*} T' \xrightarrow{\mathsf{PO}} n' \xrightarrow{\mathsf{CL}}{}^{*} n$ holds of $G$. However, as we showed before, if $T$ if followed by a transaction in $G'$, we obtain a contradiction to $(H', G') \in \mathsf{INV}_2$. ◀

▶ **Lemma 43.** *Let $(H', G')$ be the result of performing an update* TXREAD$(T, x)$ *on $(H, G)$. Assume that $(H, G), (H', G') \in \mathsf{INV}$ and $G$ is acyclic. Then $G'$ is acyclic too.*

**Proof.** By contrapositive: we assume that $G'$ contains a simple cycle and show that $G'$ violates $\mathsf{INV}$. The graph update only adds edges of the form $_ \xrightarrow{\mathsf{DEP}_x} T$ (in particular, Proposition 41 rules out a possibility of new anti-dependencies). Since all new edges end in the same vertex $T$, they cannot occur in the same simple cycle. Hence, we can consider them separately.

Consider a simple cycle involving a new edge $\nu \xrightarrow{\mathsf{DEP}_x} T$ for some vertex $\nu$. By our assumption, there must be a reverse path $T \xrightarrow{\mathsf{DEP}}{}^{*} \nu$ in $G'$. Let $\nu'$ be the first action on the path after $T$. Note that $T$ is an uncompleted transaction, since it currently performs a graph update, and whose last action in $H'$ is a read response. Thus, if $\nu'$ is some transaction $T'$, we obtain a contradiction to $\mathsf{INV}_2$. Let us now consider the case when $\nu'$ is a non-transactional access $n$. The edge $T \xrightarrow{\mathsf{DEP}}{}^{*} \nu$ must be present in $G$ as well as in $G'$, as it is not added by the current graph update TXREAD$(T, x)$. Since $G$ is acyclic and $H$ is consistent and CDRF, by Lemma 16 there exist $T'$ and $n'$ such that $T \xrightarrow{\mathsf{txDEP}}{}^{*} T' \xrightarrow{\mathsf{PO}} n' \xrightarrow{\mathsf{CL}}{}^{*} n$ holds of $G$. As graph updates only add new edges and vertexes, and never remove the existing ones, it must be the case that $T \xrightarrow{\mathsf{txDEP}}{}^{*} T'$ holds of $G'$ too. However, as we showed before, if $T$ if followed by a transaction in $G'$, we obtain a contradiction to $(H', G') \in \mathsf{INV}_2$. ◀

## F Additional Details for §8: Proof of Proposition 26

Consider a TM system that guarantees strongly atomic semantics to TDRF programs, and a TM execution $\varphi$ of a single atomic block where the latter commits. Let $(_, _, \mathsf{write}(x, v))$ be its last write request to $x$. We first prove that $\varphi$ also contains a write-back $(_, _, \mathsf{wb}(x, v))$. Note that $\varphi$ is also a prefix of an execution of a program consisting only of the same atomic block followed by a non-transactional read from $x$, which is trivially TDRF. Since the TM guarantees strongly atomic semantics to this program, the read must return $v$, and therefore, the execution of $\varphi$ must contain a write-back $(_, _, \mathsf{wb}(x, v))$.

We now prove that all write-backs to $x$ occur in $\varphi$ after the first write request to $x$. Let us assume that is not the case. Let $w = (_, _, \mathsf{wb}(x, v'))$ be the first write-back to $x$, and let $\varphi' w$ be the corresponding prefix of $\varphi$. In the following, we do a case split depending on whether $v' = v_{\mathsf{init}}$.

We first consider the case when $v = v_{\mathsf{init}}$. It is easy to see that $\varphi' w$ is also a prefix of an execution of a program that runs a modification of the original atomic block where every write to $x$ is replaced by `skip` (a primitive command that never changes the state) in parallel with a non-transactional write to $x$ of $v'' \neq v_{\mathsf{init}}$ followed by a non-transactional read from $x$:

$$\{\, \mathtt{x} = v_{\mathsf{init}} \,\}$$

$$\texttt{atomic \{ ... \}} \; /\!/ T \quad \Bigg\| \quad \begin{array}{l} \mathtt{x = } v''\mathtt{;} \; /\!/ n \\ \mathtt{l = x;} \; /\!/ n' \end{array}$$

This program is trivially TDRF. Since the TM does not impose restrictions on the placement of non-transactional accesses (§2), $\varphi' n w n'$, where $n$ is a non-transactional write of $v''$ to $x$ and $n'$ is a non-transactional read of $v_{\mathsf{init}}$ from $x$, is an execution of the above program. But this execution contradicts the fact that the TM guarantees strongly atomic semantics to this program, since the semantics requires the read to return $v''$. This contradiction shows the required.

We now return to the case when $v' \neq v_{\mathsf{init}}$. It is easy to see that $\varphi' w$ is also a prefix of an execution of a program that runs a modification of the original atomic block where every write to $x$ is replaced by `skip` in parallel with a non-transactional read from $x$. This program is trivially TDRF. The TM system cannot prevent programs from executing non-transactional operations in threads with no active transactions (§2), so the program can continue $\varphi' w$ by executing the non-transactional read from $x$, which has to return the value $v'$ written by $w$. But the resulting execution contradicts the fact that the TM guarantees strongly atomic semantics to this program, since the semantics requires the read to return $v_{\mathsf{init}} \neq v'$ . This contradiction shows the required. ◀

## G Additional Details for §9: DRF for Privatization-Unsafe TMs

Since TMs such as TL2 can be made privatization-safe by conservatively inserting fences after each transaction, one might ask whether our DRF notion could be derived from the one we proposed previously [26] by specializing it to this fence placement. However, this is not the case. The reason is that one of the relations included into the happens-before of [26] (used to handle publication) is smaller than necessary. By changing the relation to incorporate an analog of the effect order, we can obtain a notion of DRF that does specialize to TDRF when fences are placed after each transaction; the techniques from this paper can be used to strengthen the theorems of [26] to handle the improved DRF. In the following, we present our improvement of the notion of data-race freedom from [26].

We assume that the set of interface actions includes a request $(a, t, \mathsf{fbegin})$ and a response $(a, t, \mathsf{fend})$ (where $a \in \mathsf{ActionId}$ and $t \in \mathsf{ThreadID}$). The actions fbegin and fend denote the beginning and, respectively, the end of the execution of a `fence` command.

A data race happens between a pair of *conflicting* actions, as defined in §3. For such actions to form a data race, they should be *concurrent*. We formalize this using a *happens-before* relation $\mathsf{fhb}(H)$ on actions in a history $H$.

For a history $H$, we define several relations over actions of $H$, which we explain in the following:

- *restricted per-thread order* $\mathsf{xpo}(H)$: $\alpha <_{\mathsf{xpo}(H)} \alpha'$ iff $\alpha <_H \alpha'$, actions $\alpha$ and $\alpha'$ are by the same thread $t$, and there is a $(_, t, \mathsf{begintx})$ action between $\alpha$ and $\alpha'$.
- *after-fence order* $\mathsf{afs}(H)$: $\alpha <_{\mathsf{afs}(H)} \alpha'$ iff $\alpha <_H \alpha'$, $\alpha = (_, _, \mathsf{fbegin})$ and $\alpha' = (_, _, \mathsf{begintx})$, i.e., the transaction begins after the fence does (Figure 7(a)).
- *before-fence order* $\mathsf{bfe}(H)$: $\alpha <_{\mathsf{bfe}(H)} \alpha'$ iff $\alpha <_H \alpha'$, $\alpha \in \{(_, _, \mathsf{committed}), (_, _, \mathsf{aborted})\}$ and $\alpha' = (_, _, \mathsf{fend})$, i.e., the transaction ends before the fence does (Figure 7(b)).

▶ **Definition 44.** *For a history $H$ we let the* happens-before *relation of $H$ be*

$$\mathsf{fhb}(H) = (\mathsf{po}(H) \cup \mathsf{cl}(H) \cup \mathsf{afs}(H) \cup \mathsf{bfe}(H) \cup (\mathsf{xpo}(H) \mathbin{;} \mathsf{ef}(H)))^+.$$

▶ **Definition 45.** *A history $H$ is* data-race free (DRF)*, written $\mathsf{DRF}(H)$, if for every history $S \in \mathcal{H}_{\mathsf{atomic}}$ such that $H \sqsubseteq S$, $\mathsf{fhb}(S)$ relates every conflict in $H$.*

▶ **Definition 46.** *A program $P$ is* data-race free (DRF) *when executed from a state $s$ with a TM $\mathcal{H}$, written $\mathsf{DRF}(P, \mathcal{H})$, if $\forall \tau \in [\![P]\!](\mathcal{H}, s).\, \mathsf{DRF}(\mathsf{history}(\tau))$.*

We have $\mathsf{xpo}(H) \mathbin{;} \mathsf{ef}(H) \subseteq \mathsf{fhb}(H)$. Intuitively, this is because, if we have $(\alpha, \alpha') \in \mathsf{ef}(H)$, then the commands by the thread of $\alpha$ preceding the transaction of $\alpha$ are guaranteed to have taken effect by the time $\alpha'$ executes. This ensures that publication can be done safely, as we now illustrate by showing that the program in Figure 2 is DRF under $\mathcal{H}_{\mathsf{atomic}}$. Traces of the program may have only a single pair of conflicting actions—the accesses to `x` in $n$ and $T_2$. Recall that, under $\mathcal{H}_{\mathsf{atomic}}$, transactions do not interleave with other transactions or non-transactional accesses. Hence, for both conflicting actions to occur, $T_1$ should execute before $T_2$, yielding a history of the form $n T_1 T_2$. In this history, we have a effect order between the write to `x_is_private` in $T_1$ and the read from `x_is_private` in $T_2$. But then the write to `x` in $n$ happens-before the read from `x` in $T_2$, so that these actions cannot form a race.

Relations $\mathsf{afs}(H)$ and $\mathsf{bfe}(H)$ are used to formalize synchronization ensured by transactional fences. Recall that a fence blocks until all active transactions complete, by either committing or aborting. Hence, every transaction either begins after the fence does (and thus the fence does not need to wait for it; Figure 7(a)) or ends (including any required clean-up) before the fence does (Figure 7(b)). The relations $\mathsf{afs}(H)$ and $\mathsf{bfe}(H)$ capture the two respective cases. Note that, as required by the semantics of fences, every transaction has to be related to a fence at least by one of the two relations: a transaction may not span a fence.

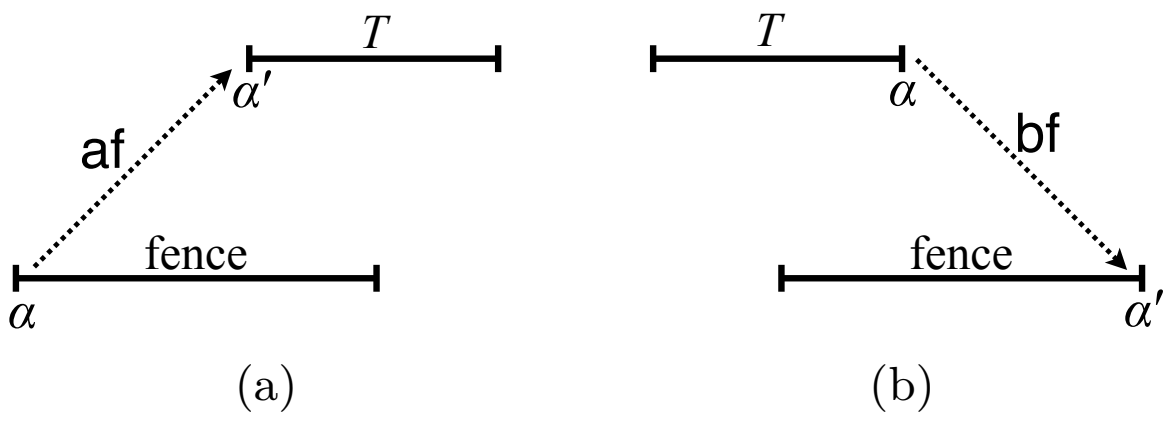

**Figure 7** An illustration of the fence relations.

Including after-fence and before-fence relations into happens-before ensures that privatization can be done safely given an appropriate placement of fences. To illustrate this, we show that the program in Figure 1 are DRF under $\mathcal{H}_{\mathsf{atomic}}$ when we place a transactional fence between $T_1$ and $n$. The possible conflicts are between the accesses to `x` in $n$ and $T_2$. For a conflict to occur, $T_2$ should execute before $T_1$, yielding a history $H$ of the form $T_2\,T_1\,\alpha_1\,\alpha_2\,n$, where $\alpha_1$ and $\alpha_2$ denote the request and the response actions of the fence. Since $T_2$ occurs before $\alpha_2$ in the history, they are related by the before-fence relation. But then the accesses to `x` in $T_2$ happen-before the write in $n$ and, therefore, the conflicting actions do not form a race. Finally, the program in Figure 3 is racy, since its traces contain pairs of conflicting actions unordered in happens-before.